\documentclass[aps,11pt]{article}
\usepackage{amssymb}
\usepackage{amsfonts}
\usepackage{amsmath,bm}
\usepackage{epsf}
\usepackage{graphicx}

\begin{document}

\title{On a paradox in the impact dynamics of smooth rigid bodies}
\author{Peter Palffy-Muhoray$^{1}$ \and Epifanio G. Virga$^{2}\thanks{
On leave from Dipartimento di Matematica, Universit\`{a} di Pavia, Pavia,
Italy}$ \and Mark Wilkinson$^{3}$ \and Xiaoyu Zheng$^{4}$ \\
$^{1}$Liquid Crystal Institute, Kent State University, OH, USA\\
$^{2}$Mathematical Institute, University of Oxford, Oxford, UK\\
$^{3}$Mathematical and Computer Sciences, Heriot-Watt University, Edinburgh,
UK\\
$^{4}$Department of Mathematical Sciences, Kent State University, OH, USA\\
}
\maketitle

\begin{abstract}
Paradoxes in the impact dynamics of rigid bodies are known to arise in the
presence of friction. We show here that, on specific occasions, in the
absence of friction, the conservation laws of classical mechanics are also
incompatible with the collisions of smooth, strictly convex rigid bodies.
Under the assumption that the impact impulse is along the normal direction
to the surface at the contact point, two convex rigid bodies which are well
separated can come into contact, and then interpenetrate each other. This
paradox can be constructed in both 2D and 3D when the collisions are
tangential, in which case no momentum or energy transfer between the two
bodies is possible. The postcollisional interpenetration can be realized
through the contact points or through neighboring points only. The
penetration distance is shown to be $O(t^{3})$. The conclusion is that rigid
body dynamics is not compatible with the conservation laws of classical
mechanics.
\end{abstract}

\section{Introduction}

As is the case with dragons \cite{Lem}, everyone knows that rigid bodies do
not exist. Nonetheless, the study of the dynamics of rigid bodies can give
valuable insights into the behavior of real compressible many-body systems.
We refer to \cite{brogliato:nonsmooth} for a recent, broad account of this
old, but still flourishing, subject.

Systems of hard spheres, interacting via steric repulsion, have also
received a great deal of attention in simulating soft matter systems. The
\textit{hard-sphere paradigm} is the assumption that, to a good
approximation, any simple liquid with strongly repulsive forces may be
modeled as a system of hard spheres. Typically, in the hard sphere paradigm,
hard sphere particles have no rotational degrees of freedom, and move in
straight lines with constant velocity until they collide elastically; that
is, by conserving linear momentum and kinetic energy. A current review is
provided in Ref.\ \cite{Dyer}. By contrast, the dynamics of systems of
non-spherical hard particles have received far less attention.

The dynamics of non-spherical particles have been addressed in a recent work
\cite{Laure}, extending the theory of the Boltzmann equation from hard
spheres to general hard particles. In Ref.\ \cite{Laure} it is suggested
that it may not always be \textquotedblleft possible to construct a family
of scattering matrices corresponding to the collision of two non-spherical
particles which conserves their total linear momentum, angular momentum and
kinetic energy\textquotedblright; that is, the dynamics of non-spherical
hard particles may not be fully compatible with the conservation laws of
classical mechanics. The specific concern was that the dynamics prescribed
by the conservation laws could, in rare instances, result in the
interpenetration of the colliding bodies.

Paradoxes in the impact dynamics of rigid bodies have been known to exist
for a long time. The classical balance laws of mechanics are not sufficient
to solve the problem of impact between two rigid bodies, that is, to predict
the motion of two colliding bodies after the impact, once their motion
before impact is known. The classical balance laws must be supplemented by
an additional \emph{impact} (or \emph{collision}) law, which is constitutive
in nature and must ultimately be justified (or at least confirmed)
experimentally.\footnote{%
Many such laws have been proposed in the past: in particular, we refer the
interested reader to \cite{chatterjee:two,chatterjee:new}, where new laws
are advanced and they are also contrasted with the vast repertoire of
pre-existing laws.} As lucidly explained in Chapter 4 of \cite%
{brogliato:nonsmooth}, impact laws fall into three broad categories. They
all relate mechanical properties of the colliding bodies before and after
impact. What distinguishes the three categories is the nature of these
properties: they may relate velocities, impulses, or kinetic energies. While
the first two categories are classical, having already been introduced in
the works of Newton and Poisson (see also \cite{glocker:frictionless}), the
third category is rather more recent in its inception: it was introduced by
Stronge~\cite{stronge:rigid} to overcome paradoxes that arise in the
presence of friction \cite{kane:dynamics,keller:impact,brach:rigid}.%
\footnote{%
Other paradoxes arise in the dynamics of colliding rigid bodies, but they
pertain more to the laws of friction than to the laws of impact. Among
these, we just mention the impact variant of Painlev\'{e} paradox \cite%
{painleve:lois,stewart:rigid-body}.} These paradoxes showed instances in
which Newton's law of impact would imply an energy gain. However, as also
remarked in \cite{stronge:rigid}, for \emph{smooth} (frictionless) rigid
bodies all three categories are equivalent and reduce to one and the same
prescription, which is consistent with the conservation of linear momentum,
angular momentum, and kinetic energy. The known paradoxes of impact dynamics
simply evaporate as friction is neglected.

In this paper, we examine collisions of smooth, strictly-convex hard
particles, and conclude that although these almost always satisfy the
conservation laws without interpenetration, there may occur rare events
where, remarkably, this is not the case. We conclude therefore that the
dynamics of strictly-convex hard particles are, in general, \emph{not}
consistent with the conservation laws of classical mechanics. Alternatively,
we could say that enforcing momentum conservation and rigidity in the
instances shown here would violate energy conservation. This is the new
paradox described in this paper.

The outline of the paper is as follows. Section $2$ on conservation laws
establishes the connection between momentum transfer and the motion of rigid
bodies. Section $3$ provides a simple 2D paradigm which illustrates the
basis of the paradox in an elementary way. Section $4$ is divided into two
parts. The first part, $4.1$, outlines how the surfaces of the colliding
bodies are characterized, and how the distances between them are measured.
The points neighboring the points of contact play an important role, and
their relevant kinematics are described. The second part, $4.2$, identifies
specific collision scenarios, and classifies their behavior. The main result
of the paper, the inconsistency of rigid body dynamics and the conservation
laws of classical mechanics is demonstrated. Section $5$ summarizes our
results. The paper is accompanied by a Supplementary Data file. In
particular, Supplementary Data I gives one exact and analytically described
example in 2D, and two numerical examples of collisions in 2D and 3D
illustrating the inconsistency. Supplementary Data II provides essential
information on the explicit definitions of terms appearing in the main body
of the paper. Supplementary Data III provides a detailed derivation of the
equations of motion, while Supplementary Data IV gives proofs of assertions
made in the main body of the paper.

\section{Conservation laws and collision dynamics}

In this section, we consider the frictionless collisions of particles which
are strictly-convex rigid bodies of arbitrary but smooth shape. We assume
that the closed bounding surfaces of the bodies are sufficiently smooth to
allow all partial derivatives up to second order to be well defined at each
point on the surface. We examine how two arbitrary but strictly-convex rigid
bodies, body $1$ and body $2$, change their momenta upon collision.

A collision is an event when particles interact and may exchange momentum.
We consider collisions where the particles are well separated before
collision, and, following their trajectories, come into a single point
contact and can interact with each other. The interaction is via a hard core
interaction potential, which is positive infinite if the particles
interpenetrate, and is zero otherwise. Since the particle energies are
finite, interpenetration of one rigid body by another, where two particles
have more than one point in common, is not possible. Thus, a collision is an
event when two particles, which were well separated before, are in single
point contact externally. Since the forces the particles exert on each other
are gradients of the potential, if forces are exerted, their magnitude is
infinite. Since the momenta and kinetic energies are finite, collisions are
necessarily instantaneous; that is, the impulse is a delta function in time.

In elastic collisions, linear and angular momenta and kinetic energy may be
exchanged, but must be conserved. Unlike compressible bodies, rigid body
systems have no mechanism for dissipation, and since there is no potential
energy except at the instant of the collision, kinetic energy must be
conserved.

We assume that the bodies are frictionless; that is, the direction of the
exchanged linear momentum is along the common normal to the two surfaces at
the point of contact. If the impulse acting on body $1$ is $-\alpha \hat{%
\mathbf{n}}_{1}$, where $\alpha $ is the magnitude of the impulse and $\hat{%
\mathbf{n}}_{1}$ is the unit outward normal at the surface of body $1$ at
the point of contact, then the conservation of linear momentum gives%
\begin{eqnarray}
m_{1}(\mathbf{v}_{1f}-\mathbf{v}_{1i}) &=&-\alpha \hat{\mathbf{n}}_{1}, \\
m_{2}(\mathbf{v}_{2f}-\mathbf{v}_{2i}) &=&\alpha \hat{\mathbf{n}}_{1},
\end{eqnarray}%
where $m_{1}$ and $m_{2}$ are the masses of bodies $1$ and $2$ with centers
of mass at $\mathbf{r}_{c1}$ and $\mathbf{r}_{c2}$, $\mathbf{v}_{1i}$ and $%
\mathbf{v}_{2i}$ are the pre-collision and $\mathbf{v}_{1f}$ and $\mathbf{v}%
_{2f}$ are the post-collision velocities of the centers of mass. The bodies
move freely in space: in the absence of collisions, linear momentum is
unchanged, and the center of mass of each body moves with constant velocity.

Conservation of angular momentum gives
\begin{eqnarray}
\mathbf{I}_{1}(\bm{\omega }_{1f}-\bm{\omega }_{1i}) &=&-\alpha \mathbf{p}%
\times \hat{\mathbf{n}}_{1}, \\
\mathbf{I}_{2}(\bm{\omega }_{2f}-\bm{\omega }_{2i}) &=&\alpha \mathbf{q}%
\times \hat{\mathbf{n}}_{1},
\end{eqnarray}%
where $\mathbf{I}_{1}$ and $\mathbf{I}_{2}$ are the moment of inertia
tensors of bodies $1$ and $2$ about their centers of mass, $\bm{\omega }%
_{1i} $ and $\bm{\omega }_{2i}$ are pre-collision and $\bm{\omega }_{1f}$
and $\bm{\omega }_{2f}$ are post-collision angular velocities. The vector $%
\mathbf{p}$ is a body-fixed vector from the center of mass of body $1$ to
the point $P$ in body $1$ which is the contact point at the instant of
collision, and $\mathbf{q}$ is a body-fixed vector from center of mass of
body $2$ to the point $Q$ in body $2$ which is the contact point at the
instant of collision. (We shall refer to the points\ $P$ and $Q$ in general
as contact points, even though they are only in contact with each other at
the instant of collision.)

We note that in the absence of collisions, angular momentum about the center
of mass is unchanged. \ However, since in general the moment of inertia is
changing due to rotation, the angular velocity also changes in time. The
angular acceleration in an inertial reference frame is
\begin{equation}
\dot{\bm{\omega}}=-\mathbf{I}^{-1}\cdot (\bm{\omega }\times (\mathbf{I}\cdot %
\bm{\omega })),
\end{equation}%
in accordance with Euler's equations \cite{Euler}.

Finally, the conservation of kinetic energy requires that%
\begin{eqnarray}
&&\frac{1}{2}m_{1}\mathbf{v}_{1i}^{2}+\frac{1}{2}m_{2}\mathbf{v}_{2i}^{2}+%
\frac{1}{2}\bm{\omega }_{1i}\cdot \mathbf{I}_{1}\cdot \bm{\omega }_{1i}+%
\frac{1}{2}\bm{\omega }_{2i}\cdot \mathbf{I}_{2}\cdot \bm{\omega }_{2i}
\notag \\
&=&\frac{1}{2}m_{1}\mathbf{v}_{1f}^{2}+\frac{1}{2}m_{2}\mathbf{v}_{2f}^{2}+%
\frac{1}{2}\bm{\omega }_{1f}\cdot \mathbf{I}_{1}\cdot \bm{\omega }_{1f}+%
\frac{1}{2}\bm{\omega }_{2f}\cdot \mathbf{I}_{2}\cdot \bm{\omega }_{2f}.
\end{eqnarray}

We note that $\mathbf{v}_{Pi}=\mathbf{v}_{1i}+\bm{\omega }_{1i}\times
\mathbf{p}$ and $\mathbf{v}_{Qi}=\mathbf{v}_{2i}+\bm{\omega }_{2i}\times
\mathbf{q}$ are the pre-collision and $\mathbf{v}_{Pf}=\mathbf{v}_{1f}+%
\bm{\omega }_{1f}\times \mathbf{p}$ and $\mathbf{v}_{Qf}=\mathbf{v}_{2f}+%
\bm{\omega }_{2f}\times \mathbf{q}$ are the post-collision velocities of the
contact points $P$ and $Q$ on bodies $1$ and $2$. Solving for $\alpha $, we
find two solutions. One solution is $\alpha =0$, corresponding to no
momentum or energy transfer. The second solution is given by%
\begin{equation}
\alpha =2\frac{(\mathbf{v}_{Pi}-\mathbf{v}_{Qi})\cdot \hat{\mathbf{n}}_{1}}{%
\hat{\mathbf{n}}_{1}\cdot \mathbf{M}\cdot \hat{\mathbf{n}}_{1}},  \label{alp}
\end{equation}%
where%
\begin{equation}
\mathbf{M}=\left( \frac{1}{m_{1}}+\frac{1}{m_{2}}\right) \mathbb{I}-\mathbf{p%
}\times \mathbf{I}_{1}^{-1}\times \mathbf{p}-\mathbf{q}\times \mathbf{I}%
_{2}^{-1}\times \mathbf{q},
\end{equation}%
and $\mathbb{I}$ is the identity tensor.\footnote{%
The product $\mathbf{a\times B\times c}$, where $\mathbf{a}$ and $\mathbf{c}$
are vectors, and $\mathbf{B}$ is a tensor, is a tensor with Cartesian
components $\varepsilon _{\alpha \beta \gamma }a_{\beta }B_{\gamma \delta
}\varepsilon _{\mu \delta \nu }c_{\nu }$, where $\varepsilon _{\alpha \beta
\gamma }$ is the Levi-Civita symbol.} The cross-products are of the vectors
and the eigenvectors of the tensors in their canonical form. Eq.\ (\ref{alp}%
) implies that the magnitude of the impulse, and of the exchanged momenta,
is proportional to the normal velocity of approach of the contact points.

The velocity of separation is given by%
\begin{equation}
\mathbf{v}_{Pf}-\mathbf{v}_{Qf}=(\mathbf{v}_{Pi}-\mathbf{v}_{Qi})\cdot \left[
\mathbb{I}-2\hat{\mathbf{n}}_{1}\hat{\mathbf{n}}_{1}\cdot \frac{\mathbf{M}}{%
\hat{\mathbf{n}}_{1}\cdot \mathbf{M}\cdot \hat{\mathbf{n}}_{1}}\right] .
\end{equation}%
It follows at once that
\begin{equation}
(\mathbf{v}_{Pf}-\mathbf{v}_{Qf})\cdot \hat{\mathbf{n}}_{1}=-(\mathbf{v}_{Pi}%
\mathbf{-\mathbf{v}}_{Qi})\cdot \hat{\mathbf{n}}_{1},  \label{rev}
\end{equation}%
in agreement with \cite{Crawford}. That is, the post-collision speed of
separation of the contact points along the normal is equal to their
pre-collision speed of approach along the normal. This result is essential
to our arguments below.

We note that if the collision were not instantaneous, then during the
collision the speed of separation would, for some instants of time, differ
from the final speed of separation \cite{Kilmister&Reeve}. This would
clearly violate energy conservation, since, unlike in the case of
interactions via soft potentials, energy cannot be stored as potential
energy. Rigid body collisions must, again, therefore be instantaneous.

In what follows, we distinguish two types of collisions.

In the first type, which we call `normal' collisions, the two particles
approach each other with a non-zero normal velocity of approach of the
contact points; here we note that $\alpha \neq 0$. In this case, there is
instantaneous momentum transfer, with the magnitude indicated in Eq.\ (\ref%
{alp}), and, in general, there is also energy transfer between the particles.

In the second type, which we call `tangential' collisions, the contact
points approach each other along the tangent to the surfaces at the points
of contact, and in this way come into contact with each other. Here $\alpha
=0$, and in this case, there is neither momentum nor energy transfer.

\section{A 2D Paradigm}

\begin{figure}[th]
\hbox{\hskip 3.5cm\includegraphics[width=9cm,
height=6cm]{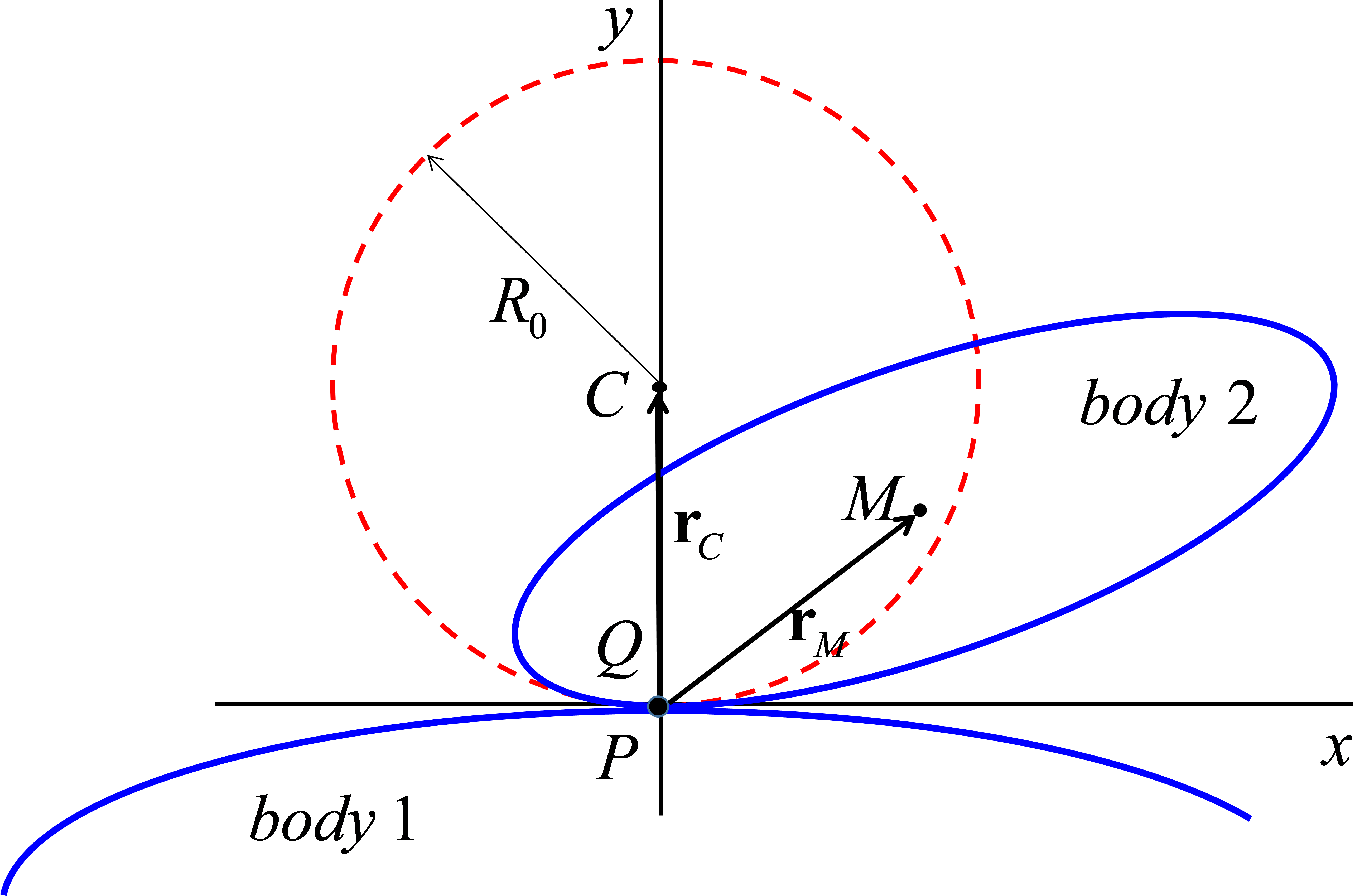}}
\caption{ Collision of two convex bodies in 2D. Body 2 with center of mass $%
M $ at $\mathbf{r}_{M}$, translating with constant velocity $\mathbf{v}_{M}$
and rotating with constant angular velocity $\bm{\omega}$ comes into contact
at the origin with body 1 at rest. The radius of curvature for body 2 at the
contact point $Q$ is $R_0$, and the center of curvature $C$ is at position $%
\mathbf{r}_{C}$. }
\label{fig-2D}
\end{figure}

To gain some initial insight into the collisions under study, we consider,
as an illustrative example, the tangential collision of an ellipse with an
irregular convex body, shown in Fig.\ \ref{fig-2D}.

The collision occurs at time $t=0$, the point of collision on body 1 is $P$
and on body 2 it is $Q$. The $x$-axis coincides with the common tangent line
at the contact points of the two bodies. For simplicity, we assume that body
1 is at rest. At time $t=0$, the center of mass $M$ of body 2 is at position
$\mathbf{r}_{M}$, moving with constant velocity $\mathbf{v}_{M}$. Body 2 is
also rotating with angular velocity $\bm{\omega }$, normal to the plane of
the bodies. The radius of curvature of body 2 at the point of contact is $%
R_{0}$, the center of curvature $C$ is at position $\mathbf{r}_{C}$. We
define $\hat{\mathbf{x}}$ and $\hat{\mathbf{y}}$ as unit vectors along the $%
x-$ and $y-$axes, respectively.

The velocity $\mathbf{v}_{Q}$ of the point of contact $Q$ is given by%
\footnote{%
For convenience, we use the 3D vector product here, although both bodies are
in 2D.}%
\begin{equation}
\mathbf{v}_{Q}=\mathbf{v}_{M}-\bm{\omega }\times \mathbf{r}_{M}.
\end{equation}%
If the collision is tangential, then $\mathbf{v}_{Q}\cdot \hat{\mathbf{y}}=0$%
, and we must have%
\begin{equation}
\mathbf{v}_{M}\cdot \hat{\mathbf{y}}=\bm{\omega }\times \mathbf{r}_{M}\cdot
\hat{\mathbf{y}},
\end{equation}%
but $\mathbf{v}_{M}\cdot \hat{\mathbf{x}}$ is arbitrary. The acceleration $%
\mathbf{a}_{Q}$ of the point of contact $Q$ is given by%
\begin{equation}
\mathbf{a}_{Q}=\omega ^{2}\mathbf{r}_{M},
\end{equation}%
and its normal acceleration is
\begin{equation}
\mathbf{a}_{Q}\cdot \hat{\mathbf{y}}=\omega ^{2}\mathbf{r}_{M}\cdot \hat{%
\mathbf{y}}.
\end{equation}

Since the relative normal velocity of the colliding bodies is zero, there is
no momentum transfer, hence body 1 remains at rest, and the motion of body 2
is unchanged after the collision.

It is interesting to consider the motion of the center of curvature $C$ and
of the circular arc near $P$ along the normal $\hat{\mathbf{y}}$. In
general, the vertical position of $C$ can be written, for small $t$, as%
\begin{equation}
\mathbf{r}_{C}(t)\cdot \hat{\mathbf{y}}=R+\mathbf{v}_{C}\cdot \hat{\mathbf{y}%
}t+\frac{1}{2}\mathbf{a}_{C}\cdot \hat{\mathbf{y}}t^{2}+\frac{1}{6}\mathbf{j}%
_{C}\cdot \hat{\mathbf{y}}t^{3}+O(t^{4}),
\end{equation}%
where $\mathbf{v}_{C}$ is the velocity, $\mathbf{a}_{C}$\thinspace is the
acceleration, and $\mathbf{j}_{C}$ is the jerk of $C$. The velocity $\mathbf{%
v}_{C}$ of $C$ along the normal is%
\begin{equation}
\mathbf{v}_{C}\cdot \hat{\mathbf{y}}=\mathbf{v}_{M}\cdot \hat{\mathbf{y}}+(%
\bm{\omega}\times (\mathbf{r}_{C}-\mathbf{r}_{M}))\cdot \hat{\mathbf{y}},
\end{equation}%
which, due to the choice of $\mathbf{v}_{M}$, is%
\begin{equation}
\mathbf{v}_{C}\cdot \hat{\mathbf{y}}=(\bm{\omega}\times \mathbf{r}_{C})\cdot
\hat{\mathbf{y}}=0.
\end{equation}%
The acceleration of $C$ along the normal is
\begin{equation}
\mathbf{a}_{C}\cdot \hat{\mathbf{y}}=-\omega ^{2}(\mathbf{r}_{C}-\mathbf{r}%
_{M})\cdot \hat{\mathbf{y}},
\end{equation}%
which vanishes in the special case when%
\begin{equation}
\mathbf{r}_{C}\cdot \hat{\mathbf{y}}=\mathbf{r}_{M}\cdot \hat{\mathbf{y}}.
\end{equation}%
In this special case, the vertical motion of $C$ is given by%
\begin{equation}
\mathbf{r}_{C}(t)\cdot \hat{\mathbf{y}}=R\hat{\mathbf{y}}+\frac{1}{6}\mathbf{%
j}_{C}\cdot \hat{\mathbf{y}}t^{3}+O(t^{4}).
\end{equation}%
The normal component of the jerk can be readily shown to be%
\begin{equation}
\mathbf{j}_{C}\cdot \hat{\mathbf{y}}=\omega ^{3}r_{M}(\hat{\mathbf{x}}\cdot
\hat{\mathbf{r}}_{M}).
\end{equation}%
We see that for a clockwise rotation (negative $\omega $), the normal
component of the center of mass velocity $\mathbf{v}_{M}$ of body 2 is
towards body 1. The normal component of the jerk $\mathbf{j}_{C}$ is
negative, indicating that the center of curvature $C$ is moving in the $-%
\hat{\mathbf{y}}$ direction, and if $|\mathbf{v}_{M}\cdot \hat{\mathbf{x}}|$
is sufficiently small, its equidistant circular arc in the vicinity of point
$P$ necessarily penetrates body 1.

This is the essence of our paradox. In certain situations, the distance
between the colliding bodies is proportional to $t^{3}$. It follows that
since the relative normal velocity is zero at the time of collision, there
is no momentum transfer, since the bodies are rigid. The subsequent motion
then leads to the interpenetration of rigid bodies -- our paradox. We also
provide, in Supplementary Data I, an example of a nonuniform disk,
corresponding to the circle shown in with dashed lines in Fig. \ref{fig-2D},
colliding with a stationary line, which allows exact analytic description of
the dynamics.

Below, we examine collisions in more detail.

\section{Description of the Dynamics}

\subsection{General Description}

Here we consider the motion of the points of contact on the colliding
bodies, as well as of the points in the vicinity of the points of contact
before and after collisions. We are interested in the compatibility of
particle motion with the conservation laws; we are particularly interested
in the separation and possible interpenetration of the two colliding
particles.


A convenient way to describe the closed surface of a body is by a
dimensionless scalar function $G(t,\mathbf{r)}$, representing level sets,
such that, if $\mathbf{r}(t)$ is the position vector of a point on the
surface of the body at time $t$, then
\begin{equation}
G(t,\mathbf{r}(t))=0.
\end{equation}%
The time $t$ appears explicitly in the argument list to indicate that the
position and the orientation of the body are, in general, changing in time.
If
\begin{equation}
\hat{\mathbf{n}}=\frac{\nabla G}{|\nabla G|}
\end{equation}%
is the outward pointing unit normal, then $G(t,\mathbf{r}(t))<0$ indicates
that the point designated by $\mathbf{r(}t\mathbf{)}$ is inside the body,
while $G(t,\mathbf{r}(t))>0$ indicates that it is outside. \ Since we are
interested, in addition, in the distances between bodies, we introduce the
signed distance
\begin{equation}
F(t,\mathbf{r}(t))=G(t,\mathbf{r}(t))L,
\end{equation}%
where $L$ is a suitably chosen length.

We now inquire whether an arbitrary point $\mathbf{r}_{2}(t)$ on the surface
of body $2$ at time $t$ is inside, outside or on the surface, described by $%
F_{1}(t,\mathbf{r}(t))=0$, of body $1$.

As we are probing collision kinematics, we are primarily interested in the
relative locations of the points of contact, but we are also interested in
the locations of material points on the surface in the neighborhood of the
contact points just before, at, and just after the collision. Due to
convexity, more distant points on the surface are also more distant from the
tangent plane at the point of contact, hence we do not consider them here.

Except perhaps at the instant of collision, particles $1$ and $2$ are moving
in space according to their force and torque-free equations of motion; that
is, with constant linear and angular momenta, conserving kinetic energy.
Their constants of motion, before and after the collision, are determined by
initial conditions. The position vector of the point of contact $P$ on body
1 is given by%
\begin{equation}
\mathbf{r}_{P}=\mathbf{r}_{c1}+\mathbf{p},
\end{equation}%
and similarly, the position vector of the point of contact $Q$ on body 2 is
given by%
\begin{equation}
\mathbf{r}_{Q}=\mathbf{r}_{c2}+\mathbf{q}.
\end{equation}%
We introduce the body fixed small vector $\bm{\varepsilon }(t)$, from the
point $Q$ to a neighboring point on body $2$. We constrain $\bm{\varepsilon }
$ to be small by requiring that $\kappa \varepsilon \ll 1$, where $\kappa $
is the maximum curvature of the normal sections of the surface at the point
of contact. The position vector of this point is
\begin{equation}
\mathbf{r}_{Q+}=\mathbf{r}_{c2}+\mathbf{q}+\bm{\varepsilon }_{Q}.
\end{equation}%
We indicate the position of a point on the surface of body $2$ in the
vicinity of $Q$ by the `$+$' sign on the subscript of $\mathbf{r}_{Q+}$. It
is useful to write $\bm{\varepsilon }_{Q}$ as
\begin{equation}
\bm{\varepsilon }_{Q}=-{\varepsilon }_{\Vert }\hat{\mathbf{n}}_{2}+%
\bm{\varepsilon }_{\bot },
\end{equation}%
where $\hat{\mathbf{n}}_{2}$ is the outward unit normal to body $2$ at $Q$, $%
\varepsilon _{\Vert }>0$ due to convexity, $\bm{\varepsilon }_{\bot }$ is
perpendicular to $\hat{\mathbf{n}}_{2}$, and $\varepsilon _{\Vert
}=O(\varepsilon _{\bot }^{2})$ as shown in Supplementary Data II.

As the bodies move, points on the surfaces of the two bodies change their
positions in space. The position $\mathbf{r}(t)$ of an arbitrary point on
the surface of a body can be expanded in Taylor series about $t=0$ to give
\begin{eqnarray}
\mathbf{r}(t) &=&\mathbf{r}(0)+\dot{\mathbf{r}}(0)t+\frac{1}{2}\ddot{\mathbf{%
r}}(0)t^{2}+\frac{1}{6}\dddot{\mathbf{r}}(0)t^{3}+O(t^{4})  \notag \\
&=&\mathbf{r}(0)+\mathbf{v}t+\frac{1}{2}\mathbf{a}t^{2}+\frac{1}{6}\mathbf{j}%
t^{3}+O(t^{4}),
\end{eqnarray}%
where $\mathbf{v}$, $\mathbf{a}$ and $\mathbf{j}$ are the velocity,
acceleration and jerk. We note here that in collisions where the relative
normal velocity is not zero, the coefficients of the powers of $t$ may
differ before and after the collision. If this is the case, we will indicate
pre-collisions values of $t$ by $t_{-}$, and post-collision values of $t$ by
$t_{+}$. Explicit expressions for these in the free motion of rigid bodies
are given in Supplementary Data III.

To determine if a point on the surface of particle $2$ with position vector $%
\mathbf{r}_{2}(t)$ has penetrated body $1$, we evaluate $F_{1}(t,\mathbf{r}%
_{2}(t))$.

It is useful to choose a specific form for $F_{1}(t,\mathbf{r}_{Q}(t))$. The
surface of a smooth convex body in the vicinity of a point on the surface
can be well described by the normal to the surface, and the two principal
curvatures, and the associated principal directions in the tangent plane. We
choose therefore $F_{1}(t,\mathbf{r)}$ such that, in the vicinity of the
point $\mathbf{r}_{P}$, one has
\begin{equation}
F_{1}(t,\mathbf{r}(t))=(\mathbf{r}-\mathbf{r}_{P})\cdot \hat{\mathbf{n}}_{1}+%
\frac{1}{2}(\mathbf{r}-\mathbf{r}_{P})\cdot (\frac{2\kappa _{1x}^{2}\hat{%
\mathbf{x}}_{1}\hat{\mathbf{x}}_{1}+2\kappa _{1y}^{2}\hat{\mathbf{y}}_{1}%
\hat{\mathbf{y}}_{1}}{|\mathbf{n}_{1}|})\cdot (\mathbf{r}-\mathbf{r}_{P})+O(|%
\mathbf{r}-\mathbf{r}_{P}|^{3}),
\end{equation}%
where $\mathbf{n}_{1}(t)=\nabla G_{1}$ is the outward normal, and we have
chosen the length $L=1/|\mathbf{n}_{1}|$. The symbols $\kappa _{1x}$ and $%
\kappa _{1y}$, both strictly positive, are the principal curvatures, and $%
\hat{\mathbf{x}}_{1}(t)$ and $\hat{\mathbf{y}}_{1}(t)$ are the corresponding
eigenvectors. The Hessian of $F_{1}$ is%
\begin{equation}
\mathbf{H}_{1}=\frac{2\kappa _{1x}^{2}\hat{\mathbf{x}}_{1}\hat{\mathbf{x}}%
_{1}+2\kappa _{1y}^{2}\hat{\mathbf{y}}_{1}\hat{\mathbf{y}}_{1}}{|\mathbf{n}%
_{1}|},
\end{equation}%
and so
\begin{equation}
F_{1}(t,\mathbf{r}(t))=(\mathbf{r}-\mathbf{r}_{P})\cdot \hat{\mathbf{n}}_{1}+%
\frac{1}{2}(\mathbf{r}-\mathbf{r}_{P})\cdot \mathbf{H}_{1}\cdot (\mathbf{r}-%
\mathbf{r}_{P})+O(|\mathbf{r}-\mathbf{r}_{P}|^{3}).
\end{equation}%
The function $F_{1}(t,\mathbf{r(}t\mathbf{))}$ measures a signed distance
between the point designated by $\mathbf{r}(t)$ and the surface of body $1$.

To indicate points in the vicinity of the point $Q$ at the time of
collision, we write
\begin{equation}
\mathbf{r}_{Q+}(t)=\mathbf{r}_{P}(t)+\bm{\delta }_{Q}(t),
\end{equation}%
where
\begin{equation}
\bm{\delta }_{Q}(t)=\mathbf{r}_{Q+}(t)-\mathbf{r}_{P}(t)=\mathbf{r}_{Q}(t)-%
\mathbf{r}_{P}(t)+\bm{\varepsilon }_{Q}(t),
\end{equation}%
We then have%
\begin{equation}
F_{1}(t,\mathbf{r}_{Q+}(t))=\hat{\mathbf{n}}_{1}\cdot \bm{\delta }_{Q}+\frac{%
1}{2}\bm{\delta }_{Q}\cdot \mathbf{H}_{1}\cdot \bm{\delta }_{Q}+O(|%
\bm{\delta }_{Q}|^{3}).
\end{equation}%
To study the approach and separation of the colliding bodies, we evaluate $%
F_{1}(t,\mathbf{r}_{Q+}(t))$ as a function of time.

The collision occurs at time $t=0$; hence we expand $F_{1}(t,\mathbf{r}%
_{Q+}(t))$ in Taylor series about $t=0$:
\begin{eqnarray}
F_{1}(t,\mathbf{r}_{Q+}(t)) &=&\mathbf{(\hat{n}}_{1}\cdot \bm{\delta }_{Q}+%
\frac{1}{2}\bm{\delta }_{Q}\cdot \mathbf{H}_{1}\cdot \bm{\delta }_{Q}){\Big |%
}_{t=0}+\frac{\partial }{\partial t}(\hat{\mathbf{n}}_{1}\cdot \bm{\delta }%
_{Q}+\frac{1}{2}\bm{\delta }_{Q}\cdot \mathbf{H}_{1}\cdot \bm{\delta }_{Q}){%
\Big |}_{t=0}t \\
&&+\frac{1}{2}\frac{\partial ^{2}}{\partial t^{2}}(\hat{\mathbf{n}}_{1}\cdot %
\bm{\delta }_{Q}+\frac{1}{2}\bm{\delta }_{Q}\cdot \mathbf{H}_{1}\cdot %
\bm{\delta }_{Q}){\Big |}_{t=0}t^{2}+O(\max (|\bm{\delta }_{Q}|^{3},\text{ }%
t^{3})).
\end{eqnarray}

Substitution gives%
\begin{eqnarray}
F_{1}(t,\mathbf{r}_{Q+}(t)) &=&(\hat{\mathbf{n}}_{1}\cdot (\mathbf{v}_{Q}-%
\mathbf{v}_{P}))t  \notag \\
&&+\frac{1}{2}(2\dot{\hat{\mathbf{n}}}_{1}\cdot (\mathbf{v}_{Q}-\mathbf{v}%
_{P})+\hat{\mathbf{n}}_{1}\cdot (\dot{\mathbf{v}}_{Q}-\dot{\mathbf{v}}_{P})+%
\mathbf{(\mathbf{v}}_{Q}-\mathbf{v}_{P})\cdot \mathbf{H}_{1}\cdot (\mathbf{v}%
_{Q}-\mathbf{v}_{P})\mathbf{)}t^{2}  \notag \\
&&+(\hat{\mathbf{n}}_{1}\cdot \bm{\varepsilon }_{Q}+\frac{1}{2}%
\bm{\varepsilon }_{Q}\cdot \mathbf{H}_{1}\cdot \bm{\varepsilon }_{Q})  \notag
\\
&&+(\dot{\hat{\mathbf{n}}}_{1}\cdot \bm{\varepsilon }_{Q}+\hat{\mathbf{n}}%
_{1}\cdot \dot{\bm{\varepsilon }}_{Q}+(\mathbf{v}_{Q}-\mathbf{v}_{P})\cdot
\mathbf{H}_{1}\cdot \mathbf{\bm{\varepsilon }}_{Q}+\dot{\bm{\varepsilon }}%
_{Q}\cdot \mathbf{H}_{1}\cdot \bm{\varepsilon }_{Q}+\frac{1}{2}%
\bm{\varepsilon }_{Q}\cdot \dot{\mathbf{H}}_{1}\cdot \bm{\varepsilon }_{Q})t
\notag \\
&&+\frac{1}{2}(\ddot{\hat{\mathbf{n}}}_{1}\cdot \bm{\varepsilon }_{Q}+2\dot{%
\hat{\mathbf{n}}}_{1}\cdot \dot{\bm{\varepsilon }}_{Q}+\hat{\mathbf{n}}%
_{1}\cdot \ddot{\bm{\varepsilon }}_{Q}  \notag \\
&&+(\dot{\mathbf{v}}_{Q}-\dot{\mathbf{v}}_{P})\cdot \mathbf{H}_{1}\cdot %
\bm{\varepsilon }_{Q}+2(\mathbf{v}_{Q}-\mathbf{v}_{P})\cdot \dot{\mathbf{H}}%
_{1}\cdot \bm{\varepsilon }_{Q}+2(\mathbf{v}_{Q}-\mathbf{v}_{P})\cdot
\mathbf{H}_{1}\cdot \dot{\bm{\varepsilon }}_{Q}  \notag \\
&&+\ddot{\bm{\varepsilon }}_{Q}\cdot \mathbf{H}_{1}\cdot \bm{\varepsilon }%
_{Q}+2\dot{\bm{\varepsilon }}_{Q}\cdot \dot{\mathbf{H}}_{1}\cdot %
\bm{\varepsilon }_{Q}+\dot{\bm{\varepsilon }}_{Q}\cdot \mathbf{H}_{1}\cdot
\dot{\bm{\varepsilon }}_{Q}+\frac{1}{2}\bm{\varepsilon }_{Q}\cdot \dot{%
\mathbf{H}}_{1}\cdot \bm{\varepsilon }_{Q})t^{2}  \notag \\
&&+O(\max (|\bm{\delta }_{Q}|^{3},\text{ }t^{3})),  \label{main}
\end{eqnarray}%
and we note that all quantities on the right hand side of Eq.\ (\ref{main})
are evaluated at $t=0$. In the case of normal collisions, the velocities $%
\mathbf{v}_{P}$ and $\mathbf{v}_{Q}$ as well as the angular velocities $%
\bm{\omega }_{1}$ and $\bm{\omega }_{2}$ change instantaneously at the
instant of collision. We therefore distinguish between pre-collision values
at $t=0^{-}$, and post-collision values at $t=0^{+}$. In the case of
tangential collisions, the velocities $\mathbf{v}_{P}$ and $\mathbf{v}_{Q}$
and the angular velocities $\bm{\omega }_{1}$ and $\bm{\omega }_{2}$ do not
change, and the distinction is not required.

We next write the expression for $F_{1}(t,\mathbf{r}_{Q+}(t))$ in standard
form,%
\begin{equation}
F_{1}(t,\mathbf{r}_{Q+}(t))=x_{Q}+v_{Q}t+\frac{1}{2}a_{Q}t^{2}+\frac{1}{6}%
j_{Q}t^{3}+O(t^{4})+f(\bm{\varepsilon }_{\bot Q},t),  \label{mot}
\end{equation}%
where
\begin{equation}
f(\bm{\varepsilon }_{\bot Q},t)=(x_{\varepsilon Q}\varepsilon _{\bot
}+x_{2\varepsilon Q}\varepsilon _{\bot }^{2})+v_{\varepsilon Q}\varepsilon
_{\bot }t+O_{\varepsilon }(\max (\varepsilon _{\bot }^{3},\varepsilon _{\bot
}^{2}t,\varepsilon _{\bot }t^{2})),  \label{mot22}
\end{equation}%
refers to the distance of a neighbor point of point $Q$, defined by $%
\bm{\varepsilon }_{\bot }$ from body $1$. Here we have introduced the symbol
$O_{\varepsilon }$ to denote `big O' for neighbor points. The index number
before the subscript $\varepsilon $ denotes the power of $\varepsilon _{\bot
}$ appearing in the expression.

Explicitly, the constants in Eqs.\ (\ref{mot}) and (\ref{mot22}) for the
point of contact $Q$ are
\begin{equation}
x_{Q}=0,
\end{equation}%
\begin{equation}
v_{Q}=\hat{\mathbf{n}}_{1}\cdot \mathbf{(v}_{Q}\mathbf{-v}_{P}\mathbf{),}
\end{equation}%
\begin{equation}
a_{Q}=\hat{\mathbf{n}}_{1}\cdot (\dot{\mathbf{v}}_{Q}-\dot{\mathbf{v}}%
_{P})+2(\bm{\omega }_{1}\times \hat{\mathbf{n}}_{1})\cdot (\mathbf{v}_{Q}-%
\mathbf{v}_{P})+(\mathbf{v}_{Q}-\mathbf{v}_{P})\cdot \mathbf{H}_{1}\cdot (%
\mathbf{v}_{Q}-\mathbf{v}_{P}),
\end{equation}%
and the additional terms for the neighboring points are%
\begin{equation}
x_{\varepsilon Q}=0,
\end{equation}%
\begin{equation}
x_{2\varepsilon Q}=\frac{1}{2}(\hat{\bm{\varepsilon}}_{\bot }\cdot \mathbf{H}%
_{2}\cdot \hat{\bm{\varepsilon}}_{\bot })+\frac{1}{2}(\hat{\bm{\varepsilon}}%
_{\bot }\cdot \mathbf{H}_{1}\cdot \hat{\bm{\varepsilon}}_{\bot }),
\end{equation}%
\begin{equation}
v_{\varepsilon Q}=\hat{\mathbf{n}}_{1}\times (\bm{\omega }_{2}-\bm{\omega }%
_{1})\cdot \hat{\bm{\varepsilon}}_{\bot }+(\mathbf{v}_{Q}-\mathbf{v}%
_{P})\cdot \mathbf{H}_{1}\cdot \hat{\bm{\varepsilon}}_{\bot },
\end{equation}%
The higher order terms have been omitted to save space, as they are not
relevant to our development below. Higher order terms in Eq.\ (\ref{mot22})
are given in Supplementary Data II for completeness.

\subsection{Analysis of Collisions}

We assume throughout that the bodies are well separated before the
collision; that is, at $t<0$, there are no shared points.

We now consider the values taken by $F_{1}(t,\mathbf{r}_{Q+}(t))$ during the
collision. We begin with the general expression%
\begin{equation}
F_{1}(t,\mathbf{r}_{Q+}(t))=v_{Q}t+\frac{1}{2}a_{Q}t^{2}+\frac{1}{6}%
j_{Q}t^{3}+O(t^{4})+f(\bm{\varepsilon }_{\bot Q},t),
\end{equation}%
where%
\begin{equation}
f(\bm{\varepsilon }_{\bot Q},t)=(x_{\varepsilon Q}\varepsilon _{\bot
}+x_{2\varepsilon Q}\varepsilon _{\bot }^{2})+v_{\varepsilon Q}\varepsilon
_{\bot }t+O_{\varepsilon }(\max (\varepsilon _{\bot }^{3},\varepsilon _{\bot
}^{2}t,\varepsilon _{\bot }t^{2})).
\end{equation}%
At the instant of collision, $f(\bm{\varepsilon }_{\bot
Q},0)=x_{2\varepsilon Q}\varepsilon _{\bot }^{2}$. Neighboring points of $Q$
on the surface of body $2$, that are also on the surface of body $1$, are
those for which $f(\bm{\varepsilon }_{\bot Q},t)=0$, that is, the point $Q$,
when $\varepsilon _{\bot }=0$, and the points for which, at the lowest order
in $t$,
\begin{equation}
\varepsilon _{\bot }=-\frac{v_{\varepsilon Q}}{x_{2\varepsilon Q}}t+O(t^{2}).
\label{eight}
\end{equation}%
Since
\begin{equation}
v_{\varepsilon Q}=[\hat{\mathbf{n}}_{1}\times (\bm{\omega }_{2}-\bm{\omega }%
_{1})+2(\mathbf{v}_{Q}-\mathbf{v}_{P})\cdot \mathbf{H}_{1}]\cdot \hat{%
\bm{\varepsilon}}_{\bot }
\end{equation}%
and
\begin{equation}
x_{2\varepsilon Q}=\frac{1}{2}\hat{\bm{\varepsilon}}_{\bot }\cdot \mathbf{H}%
_{2}\cdot \hat{\bm{\varepsilon}}_{\bot }+\frac{1}{2}\hat{\bm{\varepsilon}}%
_{\bot }\cdot \mathbf{H}_{1}\cdot \hat{\bm{\varepsilon}}_{\bot }>0,
\end{equation}%
Eq.\ (\ref{eight}) represents a closed curve on the surface of body $2$,
containing the point $Q$, and expanding in time. The shape of the curve
resembles a figure of eight: one loop corresponding to $t<0$, the other to $%
t>0$. The curve may be regarded as representing the intersection of two
bodies before, at and after collision, at small times $t$, \textit{if there
were no motion of the point} $Q$. Points on the surface of body $2$ inside
these loops are in the interior of body $1$, points outside of the loops are
outside. The points which penetrate the deepest into body $1$, are those for
which $f(\bm{\varepsilon }_{\bot Q},t)$ is at a minimum with respect to $%
\varepsilon _{\bot }$, that is, such that
\begin{equation}
\frac{\partial f(\bm{\varepsilon }_{\bot Q},t)}{\partial \varepsilon _{\bot }%
}=0.
\end{equation}%
For these points, again at the lowest order in $t$,
\begin{equation}
\varepsilon _{\bot }=-\frac{v_{\varepsilon Q}}{2x_{2\varepsilon Q}}%
t+O(t^{2}),
\end{equation}%
which represents a smaller figure of eight than that given by Eq.\ (\ref%
{eight}). For this extreme set of points, we obtain, on substitution,%
\begin{equation}
f(\bm{\varepsilon }_{\bot },t)=-\frac{v_{\varepsilon Q}^{2}}{%
4x_{2\varepsilon Q}}t^{2}+\frac{1}{6}j_{Q+}t^{3}+O_{\varepsilon }(t^{4}),
\end{equation}%
and%
\begin{equation}
F_{1}(t,\mathbf{r}_{Q+}(t))=v_{Q}t+\frac{1}{2}a_{Q}t^{2}+\frac{1}{6}%
j_{Q}t^{3}-\frac{v_{\varepsilon Q}^{2}}{4x_{2\varepsilon Q}}t^{2}+\frac{1}{6}%
j_{Q+}t^{3}+O(t^{4})+O_{\varepsilon }(t^{4}),  \label{Eq46}
\end{equation}%
and we see that the contribution of neighboring points is, to leading order,
quadratic in time.

\begin{figure}[th]
\center{\includegraphics[width=10cm,
height=6cm]{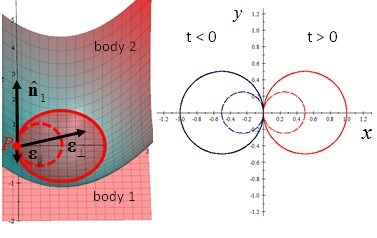}}
\caption{Figure of eight pattern illustrating penetration by neighboring
points of $Q$ that would occur if the point $Q$ was not moving. The loops of
the figure of eight need not be circles.}
\label{fig-fig8}
\end{figure}
Now we write for potentially most deeply penetrating neighboring points on
both bodies,%
\begin{eqnarray}
F_{1}(t,\mathbf{r}_{Q+}^{\ast }(t)) &=&v_{Q}t+\frac{1}{2}(a_{Q}-(\frac{%
v_{\varepsilon Q}^{2}}{2x_{2\varepsilon Q}})_{\max })t^{2}+\frac{1}{6}%
j_{Q+}^{\ast }t^{3}+O_{\varepsilon }(t^{4}), \\
F_{2}(t,\mathbf{r}_{P+}^{\ast }(t)) &=&v_{P}t+\frac{1}{2}(a_{P}-(\frac{%
v_{\varepsilon P}^{2}}{2x_{2\varepsilon P}})_{\max })t^{2}+\frac{1}{6}%
j_{P+}^{\ast }t^{3}+O_{\varepsilon }(t^{4}),
\end{eqnarray}%
where max is over all possible directions $\bm{\varepsilon}_{\bot }$, and we
have established that
\begin{eqnarray}
v_{P} &=&\hat{\mathbf{n}}_{2}\cdot (\mathbf{v}_{P}-\mathbf{v}_{Q})=v_{Q},
\label{equivalence} \\
a_{P}-(\frac{v_{\varepsilon P}^{2}}{2x_{2\varepsilon P}})_{\max } &=&a_{Q}-(%
\frac{v_{\varepsilon Q}^{2}}{2x_{2\varepsilon Q}})_{\max }.
\end{eqnarray}%
The proof of the latter is given in Supplemental Data IV. These relations
allow us to analyze the problem without bias on choice of body or of points.
We next use these results in the arguments below.

\subsubsection{Normal Collisions: $C_{-}(v<0)$}

In normal collisions, the normal speed of approach $v_{Qi}=\hat{\mathbf{n}}%
_{1}\cdot (\mathbf{v}_{Qi}-\mathbf{v}_{Pi})<0$. It changes sign during the
collision, so $v_{f}=-v_{i}$, and thus $v_{Q}t>0\,$\ both before and after
the collision. It follows that for point $Q$
\begin{equation}
F_{1}(t,\mathbf{r}_{Q}(t))=v_{Q}t+\frac{1}{2}a_{Q}t^{2}+O(t^{3})>0,
\end{equation}%
and hence, in a normal collision, the contact point $Q$ approaches body 1,
comes into contact with the point $P$, then recedes.  Convexity prevents the
neighboring points from contact with body 1, as shown by
\begin{equation}
F_{1}(t,\mathbf{r}_{Q+}^{\ast }(t))=v_{Q}t+\frac{1}{2}(a_{Q}-(\frac{%
v_{\varepsilon Q}^{2}}{2x_{2\varepsilon Q}})_{\max })t^{2}+\frac{1}{6}%
j_{Q+}^{\ast }t^{3}+O_{\varepsilon }(t^{4}).
\end{equation}
Here $j_{Q+}^{\ast }$designates the jerk for the most deeply penetrating
point in the set of $Q$ and its neighbors. \ The same argument holds for
point $P$ and its neighbors on body 1. \ This scenario can be realized in\
the 2D paradigm if $\mathbf{v}_{M}\cdot \hat{\mathbf{y}}<\bm{\omega }\times
\mathbf{r}_{M}\cdot \hat{\mathbf{y}}.$

\subsubsection{Tangential Collisions: $C_{0}(v=0)$}

In tangential collisions, the normal speed of approach $v_{Q}=\hat{\mathbf{n}%
}_{1}\cdot \mathbf{(v}_{Q}-\mathbf{v}_{P}\mathbf{)}=0$. This is an
occasional event: the normal component of the velocity difference of the
contact points must vanish at the instant of collision. As shown by Eq.\ (%
\ref{alp}), there is no momentum transfer, hence there are no changes in
either the linear or the angular velocities. The post collisional behavior
is indicated by the next terms in the expansion, as analyzed below.

\paragraph{Case C$_{0+}$}

If $a>(\frac{v_{\varepsilon }^{2}}{2x_{2\varepsilon }})_{\max }$, then for
potentially most deeply penetrating neighboring points
\begin{equation}
F_{1}(t,\mathbf{r}_{Q+}^{\ast }(t))=\frac{1}{2}(a_{Q}-(\frac{v_{\varepsilon
Q}^{2}}{2x_{2\varepsilon Q}})_{\max })t^{2}+\frac{1}{6}j_{Q+}^{\ast
}t^{3}+O_{\varepsilon }(t^{4})>0.
\end{equation}
The coefficient of the quadratic term is positive, and for small $t$, the
point $Q$ and its neighbors do not penetrate body 1. The same argument holds
for point $P$ and its neighbors on body 1. Thus there is no interpenetration
of two bodies either before or after the collision.  This case corresponds
to the 2D paradigm if $\mathbf{r}_{C}\cdot \mathbf{\hat{y}}<\mathbf{r}%
_{M}\cdot \hat{\mathbf{y}}$.

\paragraph{Case C$_{0-}$}

If $a<(\frac{v_{\varepsilon }^{2}}{2x_{2\varepsilon }})_{\max }$, then for
potentially most deeply penetrating neighboring points
\begin{equation}
F_{1}(t,\mathbf{r}_{Q+}^{\ast }\mathbf{(}t\mathbf{))}=\frac{1}{2}(a_{Q}-(%
\frac{v_{\varepsilon Q}^{2}}{2x_{2\varepsilon Q}})_{\max })t^{2}+\frac{1}{6}%
j_{Q+}^{\ast }t^{3}+O_{\varepsilon }(t^{4})<0.
\end{equation}%
For small $t$, there is penetration of body 1 by $Q$ or its neighbors before
and after the collision. This implies that the bodies were not well
separated before the collision, hence the initial conditions leading to this
collision cannot be realized. This case corresponds to the 2D paradigm if $%
\mathbf{r}_{C}\cdot \mathbf{\hat{y}}>\mathbf{r}_{M}\cdot \hat{\mathbf{y}}$.

\paragraph{Case C$_{00}$ A paradox!}

If $a=(\frac{v_{\varepsilon }^{2}}{2x_{2\varepsilon }})_{\max }$, for the
potentially most deeply penetrating neighboring points we have%
\begin{eqnarray}
F_{1}(t,\mathbf{r}_{Q+}^{\ast }(t)) &=&\frac{1}{6}j_{Q+}^{\ast
}t^{3}+O_{\varepsilon }(t^{4}), \\
F_{2}(t,\mathbf{r}_{P+}^{\ast }(t)) &=&\frac{1}{6}j_{P+}^{\ast
}t^{3}+O_{\varepsilon }(t^{4}).
\end{eqnarray}%
If the cubic terms don't vanish, then\textit{\ we face a paradox}: the two
bodies, initially well separated, come to rest at the instant of collision $%
t=0$, and begin to move and interpenetrate after the collision. This is our
key result: the dynamics of freely moving convex bodies can bring them into
contact so that the separation between them is cubic in time. At the instant
of collision the velocity of approach is zero, hence for rigid bodies there
can be no momentum exchange. The motion continues, and since the relative
acceleration is also zero, the continuing motion leads to interpenetration.
This case corresponds to the 2D paradigm if $\mathbf{r}_{C}\cdot \mathbf{%
\hat{y}=r}_{M}\cdot \hat{\mathbf{y}}$. Specific examples are provided in
Supplementary Data I.

Below, we discuss the three different types of paradoxes that can arise for
the two bodies, distinguished by the contact point behavior.

\subparagraph{Type I: $a_{P}>0$, $a_{Q}>0$}

For the distance of $P$ and $Q$ from the opposite body, we have
\begin{eqnarray}
F_{1}(t,\mathbf{r}_{Q}\mathbf{(}t\mathbf{))} &=&\frac{1}{2}a_{Q}t^{2}+\frac{1%
}{6}j_{Q}t^{3}+O(t^{4}), \\
F_{2}(t,\mathbf{r}_{P}\mathbf{(}t\mathbf{))} &=&\frac{1}{2}a_{P}t^{2}+\frac{1%
}{6}j_{P}t^{3}+O(t^{4}).
\end{eqnarray}%
Both contact points $P$ and $Q$ approach the other body, and comes into
contact, and recede from the other body. In this case, the immediate
interpenetration of two bodies is only through the immediate neighboring
points of $P$ and $Q$. $\ $This can be realized in the 2D paradigm if $%
\mathbf{v}_{M}\cdot \hat{\mathbf{x}}\neq 0$.

\subparagraph{Type II: $a_{P}=0,a_{Q}>0$}

In this case, for the distance of $P$ and $Q$ from the opposite body, we
have
\begin{eqnarray}
F_{1}(t,\mathbf{r}_{Q}(t)) &=&\frac{1}{2}a_{Q}t^{2}+\frac{1}{6}%
j_{Q}t^{3}+O(t^{4}), \\
F_{2}(t,\mathbf{r}_{P}(t)) &=&\frac{1}{6}j_{P}t^{3}+O(t^{4}).
\end{eqnarray}%
The contact point $P$ on body 1 and its immediate neighbors enter the body 2
following the collision. The contact point $Q$ on body 2 approaches, come
into contact, and recedes from body 1. The penetration into body 1 is
through immediate neighboring points of $Q$. This can be realized in the 2D
paradigm if $\mathbf{v}_{M}\cdot \hat{\mathbf{x}}=0$.

\subparagraph{Type III: $a_{P}=0,a_{Q}=0$}

In this case, for the distance of $P$ and $Q$ from the opposite body, we
have
\begin{eqnarray}
F_{1}(t,\mathbf{r}_{Q}\mathbf{(}t\mathbf{))} &=&\frac{1}{6}%
j_{Q}t^{3}+O(t^{4}), \\
F_{2}(t,\mathbf{r}_{P}\mathbf{(}t\mathbf{))} &=&\frac{1}{6}%
j_{P}t^{3}+O(t^{4}).
\end{eqnarray}%
The contact points $P$ and $Q$ together with their immediate neighbors both
enter the opposite body following the collision. This scenario cannot occur
in the 2D paradigm when one body is at rest.

We illustrate this example by the following special case. We consider the
case with the additional constraint that the relative tangential velocity is
also zero; that is, $\mathbf{v}_{Q}-\mathbf{v}_{P}=\mathbf{0}$. This only
makes the conditions $v_{Q}=a_{Q}=v_{\varepsilon Q}=0$ more easily
realizable in simulations. As before, for small $t$, $F_{1}(t,\mathbf{r}_{Q+}%
\mathbf{(}t))>0$ before the collision and $F_{1}(t,\mathbf{r}_{Q+}\mathbf{(}%
t))<0$ after the collision, indicating that initial conditions for such a
collision may be realized. With the additional constraint $\mathbf{v}_{Q}-%
\mathbf{v}_{P}=\mathbf{0}$, however, the condition $a=0$ requires that $\hat{%
\mathbf{n}}_{1}\cdot (\dot{\mathbf{v}}_{Q}-\dot{\mathbf{v}}_{P})<0$. This
cannot be achieved for convex bodies in 2D. Interestingly, for freely moving
homogeneous ellipsoids, $\mathbf{a\cdot \hat{n}}\leq 0$. The proof of this
is provided in Supplementary Data IV. Consequently, the condition $\hat{%
\mathbf{n}}_{1}\cdot (\dot{\mathbf{v}}_{Q}-\dot{\mathbf{v}}_{P})<0$ cannot
be achieved with ellipsoids. To demonstrate such a collision, showing the
penetration of the point $Q$ into body $1$, we have constructed an example
of a collision between an ellipsoid and a super-ellipsoid, where the
ellipsoid penetrates the super-ellipsoid after the collision. The details
are given in Supplementary Data I and the corresponding motion is rendered
in the accompanying movie.

The family of collisions corresponding to the various constraints are shown
in the flowchart of Fig. 3.
\begin{figure}[th]
\center{\includegraphics[width=10cm,
height=10cm]{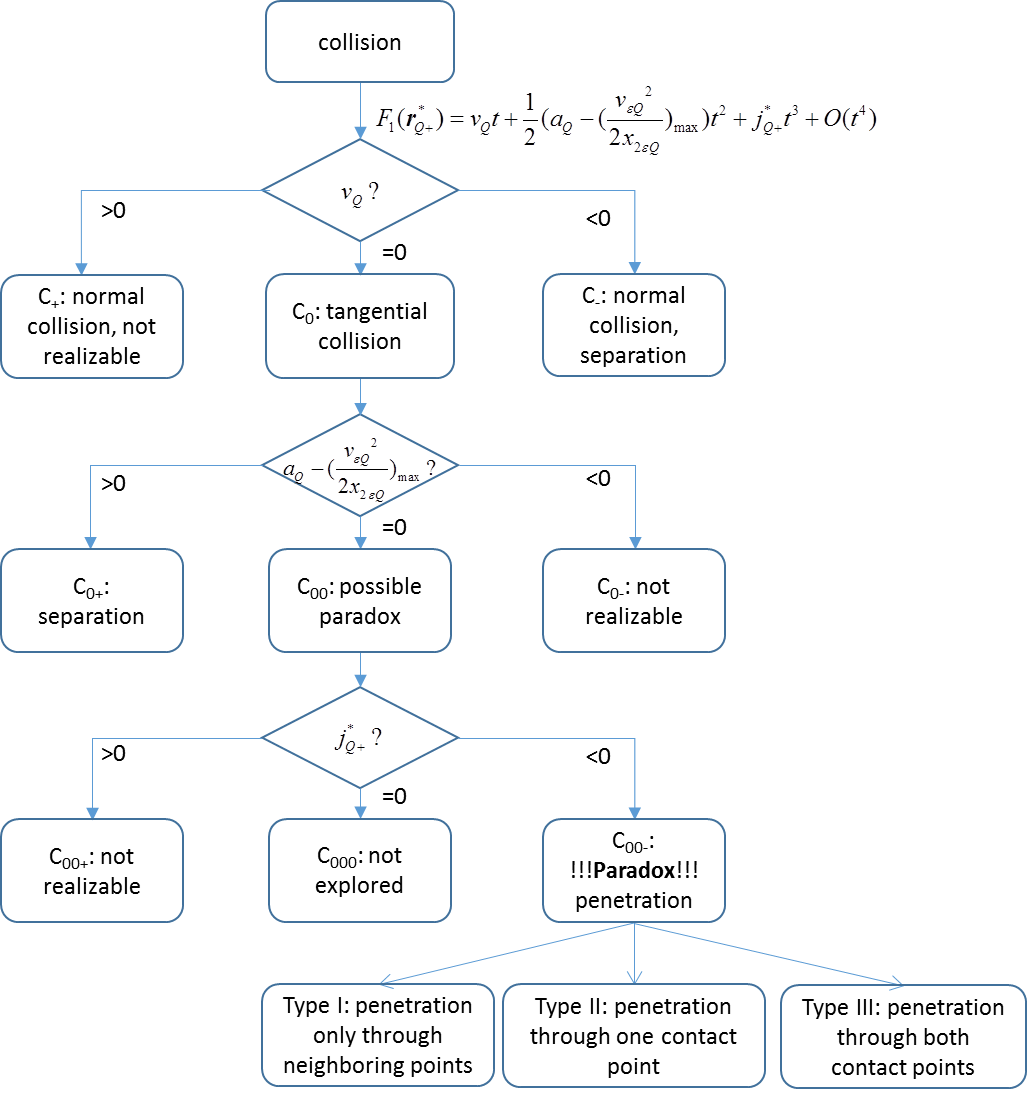}}
\caption{ A flowchart illustrating the family of collisions and indicating
where the paradox occurs.}
\end{figure}

\section{Conclusion}

We have considered above the collisions of smooth, strictly-convex rigid
bodies in 2D and 3D.

There are two types of collisions: the usual normal collisions, and the less
usual tangential collisions.

In the case of normal collisions, when the velocity of approach of two
contact points has a non-vanishing component along the normal to the surface
at the point of contact, then there will be momentum transfer between the
two bodies, and the contact points separate immediately after the collision.
Furthermore, due to the strict convexity assumption, domains of points
surrounding the contact points on two surfaces also separate immediately
after the collision. More distant points cannot penetrate the other body in
zero time, since velocities are finite. Hence there is no interpenetration
of bodies immediately after collision in the case of normal collisions.

In the case of tangential collisions, when the contact points approach each
other with a non-zero finite velocity only along the tangent to the surfaces
at the point of contact, then there is no momentum transfer, and the two
bodies pass each other without any interaction. Our analysis shows that
under certain circumstances, interpenetration occurs immediately after the
collision; this is our paradox. Simple illustrative examples are given in 2D
where one convex body collides with another at rest. We distinguish three
different types of interpenetration, characterized by the behavior of the
contact points. We have shown that the conservation of momentum and energy
lead to the interpenetration of non-spherical convex rigid bodies, which
violates their impenetrability. We conclude therefore that the dynamics of
strictly-convex rigid bodies is not consistent with the conservation laws of
classical mechanics. Specifically, the inconsistency means that the
conservation laws (conservation of linear and angular momentum, and of
energy) and rigidity (with necessarily instantaneous collisions with
infinite force and the conservation of kinetic energy) cannot hold
simultaneously in all collisions. We have enforced the conservation laws in
our approach, and observed that on some occasions rigidity is violated.
Alternatively, we could have enforced momentum conservation and rigidity,
and could have shown that, on some occasions, energy conservation is
violated, thus making our paradox for smooth rigid bodies akin to those
already known for rough rigid bodies.

The physical origins of the inconsistency are associated with the
unrealistic hard interaction, which prevents the (temporary) storage of
energy in the interaction potential. In the cases considered here, where
jerk dominates, the contact points behave as though they experience time
dependent force which is linear in time, giving rise to zero speed and
acceleration but nonzero jerk at $t=0$, followed by motion with acceleration
in the same direction. In such a situation, a soft body would respond by
gradually deforming after the collision, and temporarily storing/dissipating
energy in the deformation as a function of time. This behavior is disallowed
for rigid bodies, where kinetic energy alone must be conserved.

For non-convex rigid bodies, the collision dynamics would be much more
complicated and we have not considered it here.

In conclusion, we have presented a new paradox: as shown by our theory and
illustrated by examples, the dynamics of smooth convex rigid bodies is not
consistent with the conservation laws of classical mechanics.

\subsubsection{Data accessibility.}

All the data used in this paper are already present in the ancillary
Supplementary Data document.

\subsubsection{Authors' contributions.}

The work was undertaken by X.Z. and P. P-M., following the suggestion of
M.W. The group of X.Z., P.P-M. and M.W. was subsequently joined by E.G.V.
who enlarged the scope of the project. The paper was jointly written by all
authors.

\subsubsection{Competing interests.}

The authors have no competing interests.

\subsubsection{Funding.}

P.P-M. and X.Z. were funded for part of this work by the NSF under
DMS-1212046.

\subsubsection{Acknowledgements.}

P.P.-M. and X.Z. acknowledge support from NSF under DMS-1212046. Most of
this work was done while E.G.V. was visiting the Oxford Centre for Nonlinear
PDE at the University of Oxford, whose kind hospitality he gratefully
acknowledges. \ We are grateful to the referees for their comments and
suggestions. We are particularly grateful to Referee 2, whose insightful
comments and suggestions resulted in new results and significant
improvements to the work presented here.

\newpage

\section{Supplementary Data}

\subsection{Part I: Examples}

\subsubsection{2D Example: Collision of a nonuniform circle and a rigid wall}

Here we provide a 2D example, where the details of the collision are given
exactly in analytic form.

Consider a disk of radius $R_{0}$ with a non-homogeneous mass distribution,
such that the location of the center of mass $M$ of the disk differs from
its geometric center $C$. (This is equivalent to replacing body 1, the
homogeneous ellipse, in Fig.\ \ref{fig-2D} in the 2D Paradigm section, by an
inhomogeneous circular disk, and replacing the boundary of body 2 by a
straight line.)
\begin{figure}[th]
\center{\includegraphics[width=8cm,
height=6cm]{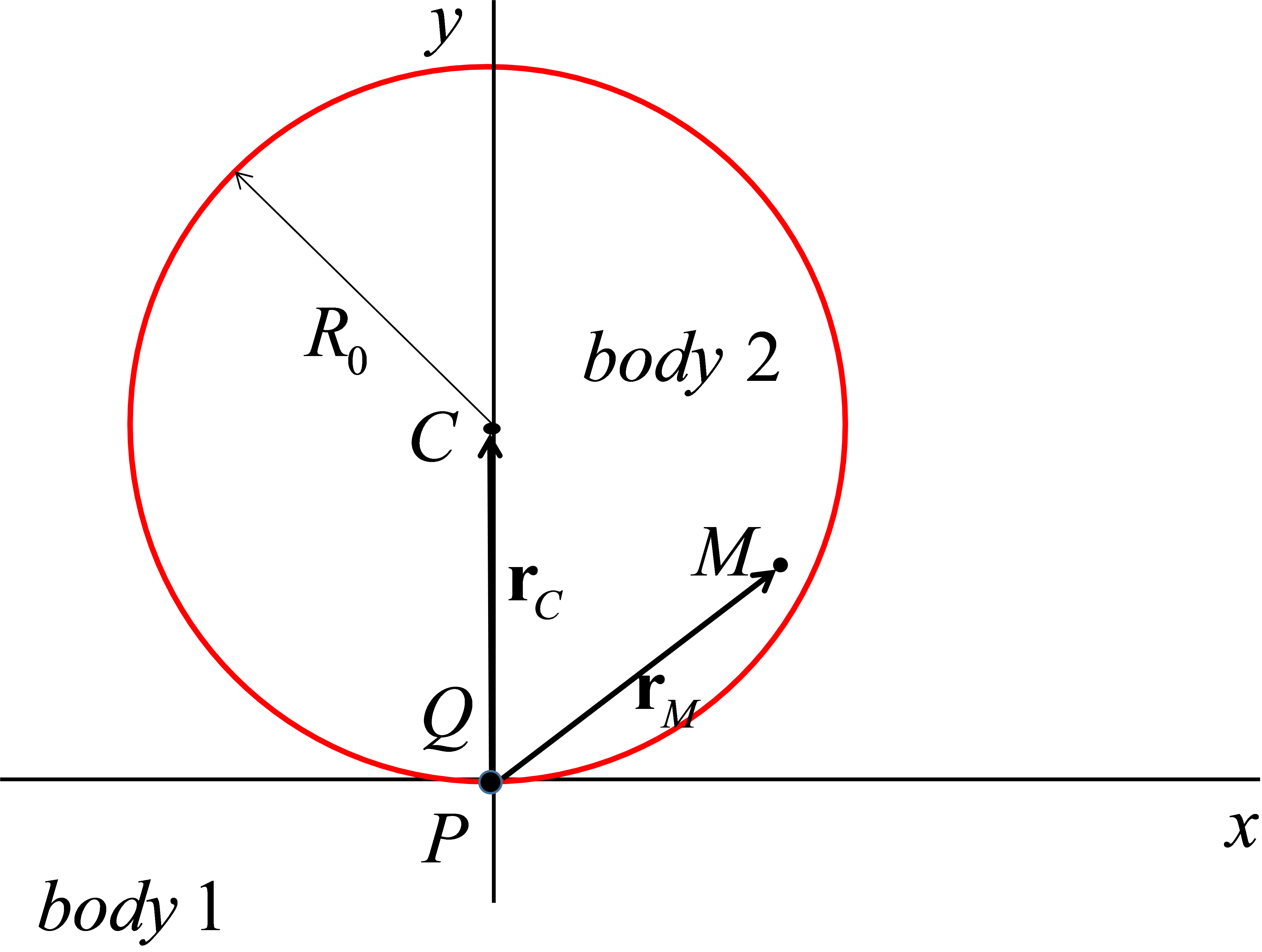}}
\caption{ A non-homogeneous disk (body 2) of radius $R_{0}$ colliding with a
wall (body 1) in an inertial frame. $M$ is the center of mass of the disk,
which differs from its geometric center $C$. $Q$ is the contact point on the
disk. $P$ is the contact point on the line, which we designate as the
origin. The contact normal of body 1 is along the $\hat{\mathbf{y}}$
direction. The disk is moving with linear velocity $\mathbf{v}_{M}$, and
rotate with angular velocity $\bm{\omega}$. }
\label{ex_disk}
\end{figure}
The moving disk collides with a line at rest along the $x$-axis in an
inertial frame. The points of contact at the instant of collision are $P$ in
body 1, and $Q$ in body 2. We denote by $\theta $ the angle that the vector $%
\mathbf{r}_{C}$ makes with $\mathbf{\hat{y}}$, and and $\theta _{M}$ is the
angle that vector $\mathbf{r}_{M}$ makes with $\mathbf{\hat{y}}$. That is,
\begin{eqnarray}
r_{Cx} &=&-R_{0}\sin \theta , \\
r_{Cy} &=&R_{0}\cos \theta ,
\end{eqnarray}%
and similarly,%
\begin{eqnarray}
r_{Mx} &=&-r_{M}\sin \theta _{M}, \\
r_{My} &=&r_{M}\cos \theta _{M}.
\end{eqnarray}%
We consider the case when the collision is tangential, that is, the normal
velocity of the point on the disk in contact with the line is zero. Hence
there is no momentum transfer, and the disk will move with constant linear
and angular velocity. We are particularly interested in the motion of points
$Q$ and $C$. Rigid body kinematics prescribes that
\begin{equation}
\mathbf{v}_{Q}=\mathbf{v}_{M}-\bm{\omega }\times \mathbf{r}_{M},
\end{equation}%
where $\mathbf{v}_{Q}$ and $\mathbf{v}_{M}$ the velocities of the points $Q$
and $M$.

Then
\begin{eqnarray}
v_{Qx} &=&v_{Mx}+\omega r_{M}\cos \theta _{M}, \\
v_{Qy} &=&v_{My}+\omega r_{M}\sin \theta _{M},
\end{eqnarray}%
and this can be integrated at once to give%
\begin{eqnarray}
Q_{x}(t) &=&v_{Mx}t+r_{M}(\sin (\theta _{M}(0)+\omega t)-\sin (\theta
_{M}(0)),  \label{qy} \\
Q_{y}(t) &=&v_{My}t-r_{M}(\cos (\theta _{M}(0)+\omega t)-\cos (\theta
_{M}(0)),
\end{eqnarray}%
where we have assumed that $Q$ is at the origin at the instant of collision $%
t=0$. The tangential collision at $t=0$ implies $v_{My}=-\omega r_{M}\sin
\theta _{M}(0)$.

Similarly, for $C$,%
\begin{equation}
\mathbf{v}_{C}=\mathbf{v}_{M}+\bm{\omega }\times (\mathbf{r}_{C}-\mathbf{r}%
_{M}),
\end{equation}%
we have%
\begin{equation}
v_{Cx}=v_{Mx}-\omega (R_{0}\cos \theta -r_{M}\cos \theta _{M}),
\end{equation}%
\begin{equation}
v_{Cy}=v_{My}-\omega (R_{0}\sin \theta -r_{M}\sin \theta _{M}).
\end{equation}%
Integration gives (on taking $\theta (0)=0$)%
\begin{eqnarray}
C_{x} &=&v_{Mx}t-R_{0}\sin \omega t+r_{M}(\sin (\theta _{M}(0)+\omega
t)-\sin (\omega _{M}(0)),  \label{cx} \\
C_{y} &=&R_{0}-\omega r_{M}\sin \theta _{M}(0)t+R_{0}(\cos \omega
t-1)-r_{M}(\cos (\theta _{M}(0)+\omega t)-\cos (\theta _{M}(0)).  \label{cy}
\end{eqnarray}%
The Eqs.\ (\ref{cx}) and (\ref{cy}) completely and exactly describe the
motion of the disk.

We now examine the motion of the bottom of the disk relative to the line.
Here only the $y-$ component matters, and we have, for small $t$,
\begin{equation}
C_{y}-R_{0}=\frac{1}{2}(r_{M}\cos \theta _{M}(0)-R_{0})\omega ^{2}t^{2}-%
\frac{1}{6}r_{M}\sin \theta _{M}(0)\omega ^{3}t^{3})+O(t^{4}).
\end{equation}

If $r_{M}\cos \theta _{M}(0)>R_{0}$, then $C_{y}-R_{0}>0$ for any small $%
t\neq 0$, which indicates that there is no interpenetration of the disk with
the wall either before or after the collision, corresponding to the case $%
C_{0+}$.

If $r_{M}\cos \theta _{M}(0)<R_{0}$, then $C_{y}-R_{0}<0$ for any small $%
t\neq 0$, indicates that there is interpenetration of the disk by the wall
both before and after the collision, corresponding to the case $C_{0-}$.

If $r_{M}\cos \theta _{M}(0)=R_{0}$, and $\omega <0$, then this leads to our
paradox $C_{00+}$.

For this 2D example, we further construct the Type I and Type II paradoxes.
We look at the motion of $Q$. Here only the $y-$ component matters, and we
have, for small $t$,

\begin{equation}
Q_{y}(t)=\frac{1}{2}R_{0}\omega ^{2}t^{2}+\frac{1}{6}d\omega
^{3}t^{3}+O(t^{4}),
\end{equation}%
where%
\begin{equation}
d=\sqrt{r_{M}^{2}-R_{0}^{2}}.
\end{equation}

Since $Q_{y}(t)>0$ for small $t$, $Q$ stays above the line; it does not
penetrate body 1$.$

We now look at the position of $P$ relative to the disk. Since%
\begin{eqnarray}
C_{x} &=&v_{Mx}t+\frac{1}{2}d\omega ^{2}t^{2}+O(t^{4}), \\
C_{y}-R_{0} &=&-\frac{1}{6}d\omega ^{3}t^{3}+O(t^{4}),
\end{eqnarray}%
the intersection of the disk with the $x$ axis, where $y=0$, occurs at the
points
\begin{equation}
x=v_{Mx}t+\frac{1}{2}d\omega ^{2}t^{2}\pm \frac{1}{\sqrt{3}}\sqrt{%
R_{0}d\omega ^{3}}t^{3/2}+O(t^{5/2}).
\end{equation}

We distinguish two cases.

If $v_{Mx}\neq 0$, at short times, the term linear in $t$ dominates, and
both intersection points are either positive or negative; they are both on
one side of $P$. Thus $P$ doesn't penetrate the disk, body 2, immediately
after the collision. In this case, the interpenetration is through
neighboring points of $P$ and neighboring points of $Q$ only, this
corresponds to our Type I paradox.

If $v_{Mx}=0$, the term of order $t^{3/2}$ dominates, and one intersection
point is positive, and the other is negative; they bracket $P$. Thus both
the point $P$ and its immediate neighbors enter the disk, but only the
points neighboring $Q$ in the disk penetrate body 1. This corresponds to our
Type II paradox.

We remark that we cannot construct the Type III paradox in this example,
where, in 2D, one of the bodies is stationary.

A brief video is provided in the following link:

http://www.math.kent.edu/\symbol{126}zheng/papers/animationDisk.gif

\subsubsection{2D Example: collision of Two Ellipses}

We note that here, and in subsequent examples, we give initial conditions
with machine precision to enable interested readers to duplicate our
simulation, and verify our results.

\noindent \indent Both ellipses have semi-axes lengths $a=2, b=1$.

The center of mass of ellipse $1$ is at $(0,0)$.

The long axis of ellipse $1$ is along $(1,0)$.

The contact point $P$ at ellipse $1$ is $%
(-1.5118578920369088,-0.6546536707079771)$.

The contact normal is $\hat{\mathbf{n}}_{1}=(-1/2,-\sqrt{3}/2)$.

The velocity of the center of mass of ellipse $1$ is $(0,0)$.

The angular velocity of ellipse $1$ is $(0,0,-1)$.

The center of mass of ellipse $2$ is at $%
(-1.4671103616817882,-1.9459065947411789)$.

The long axis of ellipse $2$ points along $(1/\sqrt{2},-1/\sqrt{2})$.

The contact point $Q$ at ellipse $2$ is $%
(-0.04474753035512054,1.2912529240332018)$.

The velocity of the center of mass of ellipse $2$ is $%
(-0.2139701987389151,0.9581598646385255)$.

The angular velocity of ellipse $2$ is $(0,0,-0.3787007446061675)$.
\begin{figure}[th]
\center{\includegraphics[width=6cm,
height=6cm]{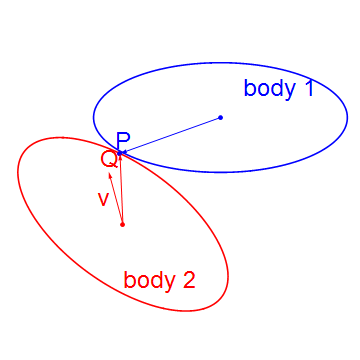}}
\caption{ The configuration of two ellipses at the time of collision. The
vector $\mathbf{v}$ emitting from the center of body 2 indicates the linear
velocity of body 2. The linear velocity of the first body is zero. Both
ellipses rotates clockwisely. The angular velocity of body 1 is $-1$, and
angular velocity of body 2 is $-0.378$. They interpenetrate each other
immediately after the collision.}
\end{figure}
This is Type II paradox, when one contact point flies away, and the other
contact point enters the opposite body. A brief animation is provided in the
following link:

http://www.math.kent.edu/\symbol{126}zheng/papers/animation2D.gif

\subsubsection{3D Example: Collision of a Superellipsoid and an Ellipsoid}

The first body is a superellipsoid described by the equation
\begin{equation*}
\frac{x^{4}}{a^{4}}+\frac{y^{4}}{b^{4}}+\frac{z^{4}}{c^{4}}=1,
\end{equation*}%
with $a=1,b=2,c=3.$

\qquad

The center of mass of superellipsoid $1$ is at $(0,0,0).$%

\qquad

The contact point $P$ at superellipsoid $1$ is

$(0.47618533191703555,1.9605250610979892,1.1955976303024505).$

\qquad

The contact normal is

$\hat{\mathbf{n}}%
_{1}=(0.22325108832916085,0.9737842723056427,0.04362502229243368).$

\qquad

The velocity of center of mass of superellipsoid $1$ is $(0,0,0).$

\qquad

The angular velocity of superellipsoid $1$ is

$(1.1740747914710616,5.121119789388775,1.2294234681418093).$%

\qquad

Ellipsoid $2$ with semiaxes length $a^{\prime }=2,b^{\prime }=c^{\prime }=1,$
and pointing along $(0,0,1),$

$(0.6239979329820173,0.781425991143224,0),$

$(-0.781425991143224,0.6239979329820173,0)$, respectively.

\qquad

The center of mass of ellipsoid $2$ is at

$(0.6988018183920255,2.931541305218898,1.3696016946556036).$%

\qquad

The contact point $Q$ at ellipsoid $2$ is

$(0.47618533191703555,1.9605250610979892,1.1955976303024505).$

\qquad

The velocity of center of mass of ellipsoid $2$ is

$(2.7414669205732727,-0.5956711297619547,-0.1367990744145808).$

\qquad

The angular velocity of ellipsoid $2$ is $(0,0,1).$


\begin{figure}[th]
\center{\includegraphics[width=6cm,
height=6cm]{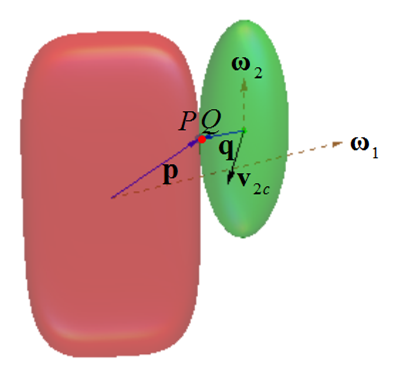}}
\caption{ The illustration shows the configuration of the superellipsoid and
ellipsoid at the time of collision. The body-fixed vectors $\mathbf{p}$ and $%
\mathbf{q}$ meet at the contact point. The vectors with dashed lines from
the centers of the two bodies represent angular velocities, and the third
solid arrow from center of the ellipsoid represents the velocity of its
center of mass. Immediately after the collision, the point $P$ and its
neighbors penetrate body $2$, the ellipsoid, and the point $Q$ and its
neighbors penetrate body $1$, the superellipsoid. A link to an animation of
the collision is provided below.}
\end{figure}

This is a type III paradox, where both contact points penetrate the opposite
body. We also note that if we let the ellipsoid spins with respect to its
long axis with any other angular velocity, then this will correspond to type
II paradox, the contact point on the ellipsoid won't penetrate the
superellipsoid. A brief animation is provided in the following link:

http://www.math.kent.edu/\symbol{126}zheng/papers/animation3D.gif

An animation of the collision between the superellipsoid and the ellipsoid
is shown on the left; the animation on the right is a blow-up of the region
near the contact point $P$ and $Q$. The penetration of the superellipsoid
into the ellipsoid can be clearly seen in the blown-up region on the right.

\subsection{Part II: Essential Information}

\subsubsection{Epsilon}

Since $\bm{\varepsilon}_{Q}$ denotes a neighboring point of $Q$ it must lie
on the surface of body $2$. It must therefore satisfy%
\begin{equation}
F_{2}(\bm{\varepsilon}_{Q})=\hat{\mathbf{n}}_{2}\cdot \bm{\varepsilon }_{Q}+%
\frac{1}{2}\bm{\varepsilon }_{Q}\cdot \mathbf{H}_{2}\cdot \bm{\varepsilon }%
_{Q}+O(|\bm{\varepsilon }_{Q}|^{3})=0.
\end{equation}%
Writing
\begin{equation}
\bm{\varepsilon }_{Q}=-\varepsilon _{\Vert }\hat{\mathbf{n}}_{2}+%
\bm{\varepsilon }_{\bot },
\end{equation}%
we have%
\begin{equation}
-\varepsilon _{\Vert }+\frac{1}{2}\mathbf{\varepsilon }_{\bot }\cdot \mathbf{%
H}_{2}\cdot \bm{\varepsilon }_{\bot }+O(|\bm{\varepsilon }_{Q}|^{3})=0,
\end{equation}%
or%
\begin{equation}
\varepsilon _{\Vert }=\varepsilon _{\bot }^{2}\frac{1}{2}\hat{%
\bm{\varepsilon}}_{\bot }\cdot \mathbf{H}_{2}\cdot \hat{\bm{\varepsilon}}%
_{\bot }+O(|\bm{\varepsilon }_{Q}|^{3}).
\end{equation}

\subsubsection{Summary of standard form}

The standard form of $F_{1}(t,\mathbf{r}_{Q+}(t))$ for small $t$ is%
\begin{eqnarray}
F_{1}(t,\mathbf{r}_{Q+}(t)) &=&x_{Q}+v_{Q}t+\frac{1}{2}a_{Q}t^{2}+\frac{1}{6}%
j_{Q}t^{3}+O(t^{4})  \notag \\
&&+(x_{\varepsilon Q}\varepsilon _{\bot }+x_{2\varepsilon Q}\varepsilon
_{\bot }^{2})+(v_{\varepsilon Q}\varepsilon _{\bot }+v_{2\varepsilon
Q}\varepsilon _{\bot }^{2}+v_{3\varepsilon Q}\varepsilon _{\bot }^{3})t
\notag \\
&&+\frac{1}{2}(a_{\varepsilon Q}\varepsilon _{\bot }+a_{2\varepsilon
Q}\varepsilon _{\bot }^{2})t^{2}+\frac{1}{6}j_{\varepsilon
Q}t^{3}+O_{\varepsilon }(t^{4}),
\end{eqnarray}%
where%
\begin{equation}
x_{Q}=0,
\end{equation}%
\begin{equation}
v_{Q}=\hat{\mathbf{n}}_{1}\cdot (\mathbf{v}_{Q}-\mathbf{v}_{P}),
\end{equation}%
\begin{equation}
a_{Q}=2(\bm{\omega }_{1}\times \hat{\mathbf{n}}_{1})\cdot (\mathbf{v}_{Q}-%
\mathbf{v}_{P})+\hat{\mathbf{n}}_{1}\cdot (\dot{\mathbf{v}}_{Q}-\dot{\mathbf{%
v}}_{P})+\mathbf{(\mathbf{v}}_{Q}-\mathbf{\mathbf{v}}_{P}\mathbf{)}\cdot
\mathbf{H}_{1}\cdot (\mathbf{v}_{Q}-\mathbf{v}_{P}),
\end{equation}%
for the point of contact $Q$, and%
\begin{equation}
x_{2\varepsilon }=\frac{1}{2}(\hat{\bm{\varepsilon}}_{\bot }\cdot \mathbf{H}%
_{1}\cdot \hat{\bm{\varepsilon}}_{\bot })+\frac{1}{2}(\hat{\bm{\varepsilon}}%
_{\bot }\cdot \mathbf{H}_{2}\cdot \hat{\bm{\varepsilon}}_{\bot }),
\end{equation}%
and
\begin{eqnarray}
v_{\varepsilon } &=&\hat{\mathbf{n}}_{1}\times (\bm{\omega }_{2}-\bm{\omega }%
_{1})\cdot \hat{\bm{\varepsilon}}_{\bot }+(\mathbf{v}_{Q}-\mathbf{v}%
_{P})\cdot \mathbf{H}_{1}\cdot \hat{\bm{\varepsilon}}_{\bot }, \\
v_{2\varepsilon } &=&(\bm{\omega }_{2}\times \hat{\bm{\varepsilon}}_{\bot
})\cdot \mathbf{H}_{1}\cdot \hat{\bm{\varepsilon}}_{\bot }+\frac{1}{2}\hat{%
\bm{\varepsilon}}_{\bot }\cdot \dot{\mathbf{H}}_{1}\cdot \hat{%
\bm{\varepsilon}}_{\bot }, \\
v_{3\varepsilon } &=&h(\hat{\bm{\varepsilon}}_{\bot })(\bm{\omega }%
_{2}\times \dot{\hat{\mathbf{n}}}_{1})\cdot \mathbf{H}_{1}\cdot \bm{
\hat{\varepsilon}}_{\bot }+h(\hat{\bm{\varepsilon}}_{\bot })\hat{\mathbf{n}}%
_{1}\cdot \dot{\mathbf{H}}_{1}\cdot \hat{\bm{\varepsilon}}_{\bot }, \\
v_{4\varepsilon } &=&\frac{1}{2}h^{2}(\hat{\bm{\varepsilon}}_{\bot })\mathbf{%
\hat{n}}_{1}\cdot \dot{\mathbf{H}}_{1}\cdot \hat{\mathbf{n}}_{1},
\end{eqnarray}%
and
\begin{eqnarray}
a_{\varepsilon } &=&(\dot{\bm{\omega }}_{1}\times \hat{\mathbf{n}}_{1})\cdot
\hat{\bm{\varepsilon}}_{\bot }+(\bm{\omega }_{1}\cdot \hat{\mathbf{n}}_{1})(%
\bm{\omega }_{1}\cdot \hat{\bm{\varepsilon}}_{\bot })+2(\hat{\bm{\varepsilon}%
}_{\bot }\cdot \hat{\mathbf{n}}_{1})(\bm{\omega }_{1}\cdot \bm{\omega }_{2})
\notag \\
&&-2(\hat{\bm{\varepsilon}}_{\bot }\cdot \bm{\omega }_{1})(\bm{\omega }%
_{2}\cdot \hat{\mathbf{n}}_{1})+\hat{\bm{\varepsilon}}_{\bot }\cdot (\mathbf{%
\hat{n}}_{1}\times \dot{\bm{\omega }}_{2})+(\hat{\mathbf{n}}_{1}\cdot %
\bm{\omega }_{2})(\bm{\omega }_{2}\cdot \hat{\bm{\varepsilon}}_{\bot })
\notag \\
&&+(\dot{\mathbf{v}}_{Q}-\dot{\mathbf{v}}_{P})\cdot \mathbf{H}_{1}\cdot \hat{%
\bm{\varepsilon}}_{\bot }+2(\mathbf{v}_{Q}-\mathbf{v}_{P})\cdot \dot{\mathbf{%
H}}_{1}\cdot \hat{\bm{\varepsilon}}_{\bot }+2(\mathbf{v}_{Q}-\mathbf{v}%
_{P})\cdot \mathbf{H}_{1}\cdot (\bm{\omega }_{2}\times \hat{\bm{\varepsilon}}%
_{\bot }\mathbf{),} \\
a_{2\varepsilon } &=&-h(\hat{\bm{\varepsilon}}_{\bot })\bm{\omega }_{1}\cdot
(\mathbb{I}-\hat{\mathbf{n}}_{1}\hat{\mathbf{n}}_{1})\cdot \bm{\omega }%
_{1}+2h(\hat{\bm{\varepsilon}}_{\bot })\bm{\omega }_{1}\cdot (\mathbb{I}-%
\hat{\mathbf{n}}_{1}\hat{\mathbf{n}}_{1})\cdot \bm{\omega }_{2}  \notag \\
&&-h(\hat{\bm{\varepsilon}}_{\bot })\bm{\omega }_{2}\cdot (\mathbb{I}-\hat{%
\mathbf{n}}_{1}\hat{\mathbf{n}}_{1})\cdot \bm{\omega }_{2}+2h(\hat{%
\bm{\varepsilon}}_{\bot })(\mathbf{v}_{Q}-\mathbf{v}_{P})\cdot \dot{\mathbf{H%
}}_{1}\cdot \hat{\mathbf{n}}_{1}  \notag \\
&&+2h(\hat{\bm{\varepsilon}}_{\bot })(\mathbf{v}_{Q}-\mathbf{v}_{P})\cdot
\mathbf{H}_{1}\cdot (\bm{\omega }_{2}\times \hat{\mathbf{n}}_{1})+(\dot{%
\bm{\omega }}_{2}\times \hat{\bm{\varepsilon}}_{\bot })\cdot \mathbf{H}%
_{1}\cdot \hat{\bm{\varepsilon}}_{\bot }  \notag \\
&&-\hat{\bm{\varepsilon}}_{\bot }\cdot \lbrack (\omega _{2}^{2}\mathbb{I}-%
\bm{\omega }_{2}\bm{\omega }_{2})\cdot \mathbf{H}_{1}]\cdot \hat{%
\bm{\varepsilon}}_{\bot }+2(\bm{\omega }_{2}\times \hat{\bm{\varepsilon}}%
_{\bot })\cdot \dot{\mathbf{H}}_{1}\cdot \hat{\bm{\varepsilon}}_{\bot }
\notag \\
&&+(\bm{\omega }_{2}\times \hat{\bm{\varepsilon}}_{\bot })\cdot \mathbf{H}%
_{1}\cdot (\bm{\omega }_{2}\times \hat{\bm{\varepsilon}}_{\bot })+h(\hat{%
\bm{\varepsilon}_{\bot }})\hat{\mathbf{n}}_{1}\cdot \ddot{\mathbf{H}}%
_{1}\cdot \hat{\bm{\varepsilon}}_{\bot }+\frac{1}{2}\hat{\bm{\varepsilon}}%
_{\bot }\cdot \ddot{\mathbf{H}}_{1}\cdot \hat{\bm{\varepsilon}}_{\bot }, \\
a_{3\varepsilon } &=&h(\hat{\bm{\varepsilon}}_{\bot })(\dot{\bm{\omega }}%
_{2}\times \hat{\mathbf{n}}_{1})+h(\hat{\bm{\varepsilon}}_{\bot })(%
\bm{\omega }_{2}\cdot \hat{\mathbf{n}}_{1})\bm{\omega }_{2}\cdot \mathbf{%
\mathbf{H}}_{1}\cdot \hat{\bm{\varepsilon}}_{\bot }  \notag \\
&&+2h(\hat{\bm{\varepsilon}}_{\bot })(\bm{\omega }_{2}\times \hat{%
\bm{\varepsilon}}_{\bot })\cdot \dot{\mathbf{H}}_{1}\cdot \hat{\mathbf{n}}%
_{1}+2h(\hat{\bm{\varepsilon}}_{\bot })(\bm{\omega }_{2}\times \mathbf{\hat{n%
}}_{1})\cdot \dot{\mathbf{H}}_{1}\cdot \hat{\bm{\varepsilon}}_{\bot }  \notag
\\
&&+h(\hat{\bm{\varepsilon}}_{\bot })\hat{\mathbf{n}}_{1}\cdot \ddot{\mathbf{H%
}}_{1}\cdot \hat{\bm{\varepsilon}}_{\bot }, \\
a_{4\varepsilon } &=&2h^{2}(\hat{\bm{\varepsilon}}_{\bot })(\bm{\omega }%
_{2}\times \hat{\mathbf{n}}_{1})\cdot \dot{\mathbf{H}}_{1}\cdot \hat{\mathbf{%
n}}_{1}+h^{2}(\hat{\bm{\varepsilon}}_{\bot })(\bm{\omega }_{2}\times \hat{%
\mathbf{n}}_{1})\cdot \mathbf{H}_{1}\cdot (\bm{\omega }_{2}\times \hat{%
\mathbf{n}}_{1})  \notag \\
&&+\frac{1}{2}h^{2}(\hat{\bm{\varepsilon}}_{\bot })\hat{\mathbf{n}}_{1}\cdot
\ddot{\mathbf{H}}_{1}\cdot \hat{\mathbf{n}}_{1},
\end{eqnarray}%
for the neighboring points.

\begin{equation}
h(\hat{\bm{\varepsilon}}_{\bot })=\frac{1}{2}\hat{\bm{\varepsilon}}_{\bot
}\cdot\mathbf{H}_{2}\cdot \hat{\bm{\varepsilon}}_{\bot }.
\end{equation}

\begin{figure}[th]
\center{\includegraphics[width=10cm,
height=10cm]{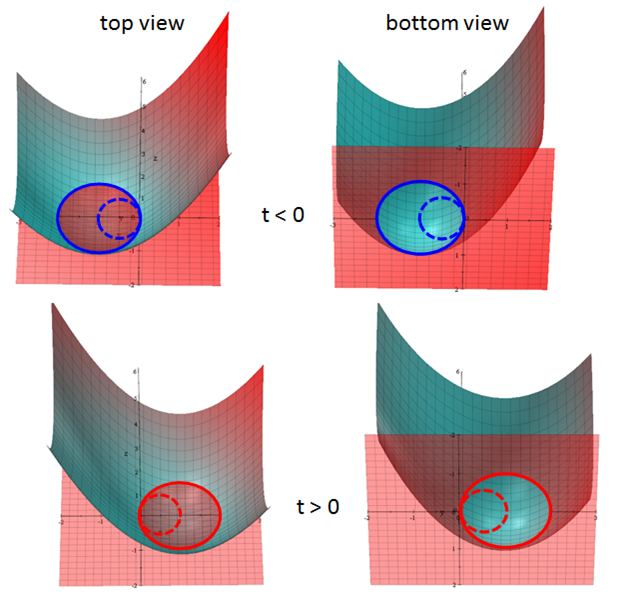}}
\caption{ Illustration of the Figure of eight pattern from different
perspectives and at different times. }
\label{fig8supp}
\end{figure}

\subsection{Part III - Information to Satisfy Completeness}

\subsection{Describing the motion}

Here we look at the coefficients in the Taylor series expansion of the
positions of points on the two bodies as functions of time.

The bodies are moving with constant linear and angular momentum, conserving
kinetic energy. If the position of a point on the surface of one body is $%
\mathbf{r}(t)$, then we have for the instantaneous velocity%
\begin{eqnarray}
\dot{\mathbf{r}} &=&\dot{\mathbf{r}}_{c}+\bm{\omega }\times (\mathbf{r}-%
\mathbf{r}_{c})  \notag \\
&=&\dot{\mathbf{r}}_{c}+\bm{\omega }\times \bm{\rho },
\end{eqnarray}%
where $\bm{\rho }=\mathbf{r}-\mathbf{r}_{c}$ is a body fixed vector, from
the center of mass of the particle to the point in question on the body.
Continuing, we have for the instantaneous acceleration%
\begin{eqnarray}
\ddot{\mathbf{r}} &=&\dot{\bm{\omega }}\times \bm{\rho }+\bm{\omega }\times
\dot{\bm{\rho }}  \notag \\
&=&\dot{\bm{\omega }}\times \bm{\rho }+\bm{\omega }\times \bm{(\omega }%
\times \bm{\rho )}  \notag \\
&=&\dot{\bm{\omega }}\times \bm{\rho }+(\bm{\omega }\cdot \bm{\rho })%
\bm{\omega }-\omega ^{2}\bm{\rho }.
\end{eqnarray}%
We note that $\ddot{\mathbf{r}}_{c}=0$ due to linear momentum conservation.
Angular momentum conservation gives (Euler's equations in the lab frame),%
\begin{equation}
\dot{\bm{\omega }}=\mathbf{-I}^{-1}\cdot (\bm{\omega }\times \mathbf{I}\cdot %
\bm{\omega }),  \label{80}
\end{equation}%
where $\mathbf{I}$ is the moment of inertia tensor, and\footnote{%
In Cartesian components, the rhs of Eq.\ (\ref{81}) would read $\varepsilon
_{\alpha \beta \gamma }\omega _{\beta }I_{\gamma \delta }-I_{\alpha \beta
}\omega _{\gamma }\varepsilon _{\delta \beta \gamma }$.}%
\begin{equation}
\dot{\mathbf{I}}=\bm{\omega }\times \mathbf{I}-\mathbf{I}\times \bm{\omega }.
\label{81}
\end{equation}

Substitution gives%
\begin{equation}
\ddot{\mathbf{r}}=-\mathbf{I}^{-1}\cdot (\bm{\omega }\times \mathbf{I}\cdot %
\bm{\omega })\times \bm{\rho }+\bm{\omega}(\bm{\omega }\cdot \bm{\rho}
)-\omega ^{2}\bm{\rho},
\end{equation}%
and
\begin{equation}
(\mathbf{I}^{-1})^{\cdot }=-\mathbf{I}^{-1}\dot{\mathbf{I}}\mathbf{I}^{-1},
\label{83}
\end{equation}%
or%
\begin{equation}
(\mathbf{I}^{-1})^{\cdot}=\bm{\omega }\times \mathbf{I}^{-1}-\mathbf{I}%
^{-1}\times \bm{\omega}.
\end{equation}

Continuing gives the instantaneous jerk%
\begin{eqnarray}
\dddot{\mathbf{r}} &=&\ddot{\bm{\omega }}\times \bm{\rho }+2\dot{\bm{\omega }%
}\times \dot{\bm{\rho }}+\bm{\omega }\times \ddot{\bm{\rho }}  \notag \\
&=&\ddot{\bm{\omega }}\times \bm{\rho }+2(\dot{\bm{\omega }}\cdot \bm{\rho })%
\bm{\omega }-3(\dot{\bm{\omega }}\cdot \bm{\omega })\bm{\rho }+(\bm{\omega }%
\cdot \bm{\rho })\dot{\bm{\omega }}-\omega ^{2}\bm{\omega }\times \bm{\rho },
\label{jerk}
\end{eqnarray}%
where, by Eq.\ (\ref{80}),%
\begin{equation}
\ddot{\bm{\omega }}=-(\mathbf{I}^{-1})^{\cdot }\cdot (\bm{\omega }\times
\mathbf{I}\cdot \bm{\omega })-\mathbf{I}^{-1}\cdot (\dot{\bm{\omega }}\times
\mathbf{I}\cdot \bm{\omega })-\mathbf{I}^{-1}\cdot (\bm{\omega }\times \dot{%
\mathbf{I}}\cdot \bm{\omega })-\mathbf{I}^{-1}\cdot (\bm{\omega }\times
\mathbf{I}\cdot \dot{\bm{\omega }}),
\end{equation}%
and $\dot{\bm{\omega }}$ and $\dot{\mathbf{I}}$ are given by Eqs.\ (\ref{80}%
) and (\ref{81}) respectively. By substitution into Eq.\ (\ref{jerk}), we
have an explicit expression for the jerk.

\subsubsection{Kinetic Equations}

The main equation can be derived at once from first principles.%
\begin{eqnarray}
F_{1}(t,\mathbf{r}_{Q+}(t)) &=&(\hat{\mathbf{n}}_{1}\cdot \bm{\delta }_{Q}+%
\frac{1}{2}\bm{\delta }_{Q}\cdot \mathbf{H}_{1}\cdot \bm{\delta }_{Q}){\Big |%
}_{t=0}  \notag \\
&&+\frac{\partial }{\partial t}(\hat{\mathbf{n}}_{1}\cdot \bm{\delta }_{Q}+%
\frac{1}{2}\bm{\delta }_{Q}\cdot \mathbf{H}_{1}\cdot \bm{\delta }_{Q}){\Big |%
}_{t=0}t  \notag \\
&&+\frac{1}{2}\frac{\partial ^{2}}{\partial t^{2}}(\hat{\mathbf{n}}_{1}\cdot %
\bm{\delta }_{Q}+\frac{1}{2}\bm{\delta }_{Q}\cdot \mathbf{H}_{1}\cdot %
\bm{\delta }_{Q}){\Big |}_{t=0}t^{2}+O(\max (|\bm{\delta }_{Q}|^{3},\text{ }%
t^{3})).
\end{eqnarray}%
Then%
\begin{eqnarray}
F_{1}(t,\mathbf{r}_{Q+}(t)) &=&(\hat{\mathbf{n}}_{1}\cdot \bm{\delta }_{Q}+%
\frac{1}{2}\bm{\delta }_{Q}\cdot \mathbf{H}_{1}\cdot \bm{\delta }_{Q}){\Big |%
}_{t=0}  \notag \\
&&+(\dot{\hat{\mathbf{n}}}_{1}\cdot \bm{\delta }_{Q}+\hat{\mathbf{n}}%
_{1}\cdot \dot{\bm{\delta }}_{Q}+\dot{\bm{\delta }}_{Q}\cdot \mathbf{H}%
_{1}\cdot \bm{\delta }_{Q}+\frac{1}{2}\bm{\delta }_{Q}\cdot \dot{\mathbf{H}}%
_{1}\cdot \bm{\delta }_{Q}){\Big |}_{t=0}t  \notag \\
&&+\frac{1}{2}(\ddot{\hat{\mathbf{n}}}_{1}\cdot \bm{\delta }_{Q}+2\dot{\hat{%
\mathbf{n}}}_{1}\cdot \dot{\bm{\delta }}_{Q}+\hat{\mathbf{n}}_{1}\cdot \ddot{%
\bm{\delta }}_{Q}  \notag \\
&&+\ddot{\bm{\delta }}_{Q}\cdot \mathbf{H}_{1}\cdot \bm{\delta }_{Q}+2\dot{%
\bm{\delta }}_{Q}\cdot \dot{\mathbf{H}}_{1}\cdot \bm{\delta }_{Q}+\dot{%
\bm{\delta }}_{Q}\cdot \mathbf{H}_{1}\cdot \dot{\bm{\delta }}_{Q}  \notag \\
&&+\frac{1}{2}\bm{\delta }_{Q}\cdot \ddot{\mathbf{H}}_{1}\cdot \bm{\delta
)}{\Big |}_{Qt=0}t^{2}+O(\max (|\bm{\delta }_{Q}|^{3},\text{ }t^{3})).
\end{eqnarray}%
Now%
\begin{equation}
\bm{\delta }_{Q}=\mathbf{(r}_{Q}-\mathbf{r}_{P})+\bm{\varepsilon }_{Q},
\end{equation}%
and substitution gives%
\begin{eqnarray}
F_{1} &=&(\hat{\mathbf{n}}_{1}\cdot (\mathbf{v}_{Q}-\mathbf{v}_{P})\mathbf{)}%
t  \notag \\
&&+\frac{1}{2}(2\dot{\hat{\mathbf{n}}}_{1}\cdot (\mathbf{v}_{Q}-\mathbf{v}%
_{P})+\hat{\mathbf{n}}_{1}\cdot (\dot{\mathbf{v}}_{Q}-\dot{\mathbf{v}}_{P})+(%
\mathbf{v}_{Q}-\mathbf{v}_{P})\cdot \mathbf{H}_{1}\cdot (\mathbf{v}_{Q}-%
\mathbf{v}_{P})\mathbf{)}t^{2}  \notag \\
&&+((\mathbf{v}_{Q}-\mathbf{v}_{P})\cdot \mathbf{H}_{1}\cdot \bm{\varepsilon
}_{Q})t  \notag \\
&&+\frac{1}{2}((\dot{\mathbf{v}}_{Q}-\dot{\mathbf{v}}_{P})\cdot \mathbf{H}%
_{1}\cdot \bm{\varepsilon }_{Q}+2(\mathbf{v}_{Q}-\mathbf{v}_{P})\cdot \dot{%
\mathbf{H}}_{1}\cdot \bm{\varepsilon }_{Q}+2(\mathbf{v}_{Q}-\mathbf{v}%
_{P})\cdot \mathbf{H}_{1}\cdot \dot{\bm{\varepsilon }}_{Q})t^{2}  \notag \\
&&+(\hat{\mathbf{n}}_{1}\cdot \bm{\varepsilon }_{Q}+\frac{1}{2}%
\bm{\varepsilon }_{Q}\cdot \mathbf{H}_{1}\cdot \bm{\varepsilon }_{Q})  \notag
\\
&&+(\dot{\hat{\mathbf{n}}}_{1}\cdot \bm{\varepsilon }_{Q}+\hat{\mathbf{n}}%
_{1}\cdot \dot{\bm{\varepsilon }}_{Q}+\dot{\bm{\varepsilon }}_{Q}\cdot
\mathbf{H}_{1}\cdot \bm{\varepsilon }_{Q}+\frac{1}{2}\bm{\varepsilon }%
_{Q}\cdot \dot{\mathbf{H}}_{1}\cdot \bm{\varepsilon }_{Q})t  \notag \\
&&+\frac{1}{2}(\ddot{\hat{\mathbf{n}}}_{1}\cdot \bm{\varepsilon }_{Q}+2\dot{%
\hat{\mathbf{n}}}_{1}\cdot \dot{\bm{\varepsilon }}_{Q}+\hat{\mathbf{n}}%
_{1}\cdot \ddot{\bm{\varepsilon }}_{Q}  \notag \\
&&+\ddot{\bm{\varepsilon }}_{Q}\cdot \mathbf{H}_{1}\cdot \bm{\varepsilon }%
_{Q}+2\dot{\bm{\varepsilon }}_{Q}\cdot \dot{\mathbf{H}}_{1}\cdot %
\bm{\varepsilon }_{Q}+\dot{\bm{\varepsilon }}_{Q}\cdot \mathbf{H}_{1}\cdot
\dot{\bm{\varepsilon }}_{Q}+\frac{1}{2}\bm{\varepsilon }_{Q}\cdot \mathbf{%
\ddot{H}}_{1}\cdot \bm{\varepsilon}_{Q})t^{2}+O(\max (|\bm{\delta }_{Q}|^{3},%
\text{ }t^{3})),
\end{eqnarray}%
or%
\begin{eqnarray}
F_{1} &=&(\hat{\mathbf{n}}_{1}\cdot (\mathbf{v}_{Q}-\mathbf{v}_{P})\mathbf{)}%
{\Big |}_{t=0}t  \notag \\
&&+\frac{1}{2}(2\dot{\hat{\mathbf{n}}}_{1}\cdot (\mathbf{v}_{Q}-\mathbf{v}%
_{P})+\hat{\mathbf{n}}_{1}\cdot (\dot{\mathbf{v}}_{Q}-\dot{\mathbf{v}}_{P})
\notag \\
&&+(\mathbf{v}_{Q}-\mathbf{v}_{P})\cdot \mathbf{H}_{1}\cdot (\mathbf{v}_{Q}-%
\mathbf{v}_{P})\mathbf{)}{\Big |}_{t=0}t^{2}  \notag \\
&&+(\hat{\mathbf{n}}_{1}\cdot \bm{\varepsilon }_{Q}+\frac{1}{2}%
\bm{\varepsilon }\cdot \mathbf{H}_{1}\cdot \bm{\varepsilon }_{Q}){\Big |}%
_{t=0}  \notag \\
&&+(\dot{\hat{\mathbf{n}}}_{1}\cdot \bm{\varepsilon }_{Q}+\hat{\mathbf{n}}%
_{1}\cdot \dot{\bm{\varepsilon }}_{Q}+(\mathbf{v}_{Q}-\mathbf{v}_{P})\cdot
\mathbf{H}_{1}\cdot \bm{\varepsilon }_{Q})+\dot{\bm{\varepsilon }}_{Q}\cdot
\mathbf{H}_{1}\cdot \bm{\varepsilon }_{Q}+\frac{1}{2}\bm{\varepsilon }%
_{Q}\cdot \dot{\mathbf{H}}_{1}\cdot \bm{\varepsilon }_{Q}){\Big |}_{t=0}t
\notag \\
&&+\frac{1}{2}(\ddot{\hat{\mathbf{n}}}_{1}\cdot \bm{\varepsilon }_{Q}+2\dot{%
\hat{\mathbf{n}}}_{1}\cdot \dot{\bm{\varepsilon }}_{Q}+\hat{\mathbf{n}}%
_{1}\cdot \ddot{\bm{\varepsilon }}_{Q}+  \notag \\
&&+(\dot{\mathbf{v}}_{Q}-\dot{\mathbf{v}}_{P})\cdot \mathbf{H}_{1}\cdot %
\bm{\varepsilon }_{Q}+2(\mathbf{v}_{Q}-\mathbf{v}_{P})\cdot \dot{\mathbf{H}}%
_{1}\cdot \bm{\varepsilon }_{Q}+2(\mathbf{v}_{Q}-\mathbf{v}_{P})\cdot
\mathbf{H}_{1}\cdot \dot{\bm{\varepsilon }}_{Q}  \notag \\
&&+\ddot{\bm{\varepsilon }}_{Q}\cdot \mathbf{H}_{1}\cdot \bm{\varepsilon }%
_{Q}+2\dot{\bm{\varepsilon }}_{Q}\cdot \dot{\mathbf{H}}_{1}\cdot \mathbf{%
\varepsilon }+\dot{\bm{\varepsilon }}\cdot \mathbf{H}_{1}\cdot \dot{%
\bm{\varepsilon }}_{Q}+\frac{1}{2}\bm{\varepsilon }_{Q}\cdot \mathbf{\ddot{H}%
}_{1}\cdot \bm{\varepsilon}_{Q}){\Big |}_{t=0}t^{2}+O(\max (|\bm{\delta }%
_{Q}|^{3},\text{ }t^{3})).
\end{eqnarray}

\subsection{Part IV: Proofs of selected claims}

\subsubsection{Proof of $a_{P}-(\frac{v_{\protect\varepsilon P}^{2}}{2x_{2%
\protect\varepsilon P}})_{\max }=a_{Q}-(\frac{v_{\protect\varepsilon Q}^{2}}{%
2x_{2\protect\varepsilon Q}})_{\max }$}

From previous analysis, for potentially most deeply penetrating neighboring
point on body 2, we have%
\begin{equation}
F_{1}(\mathbf{r}_{Q+}(t),t)=v_{Q}t+\frac{1}{2}(a_{Q}-\frac{v_{\varepsilon
Q}^{2}}{x_{2\varepsilon Q}})t^{2}+O_{\varepsilon }(t^{3}),
\end{equation}%
where%
\begin{eqnarray}
v_{Q} &=&\hat{\mathbf{n}}_{2}\cdot (\mathbf{v}_{Q}-\mathbf{v}_{P}), \\
a_{Q} &=&2(\bm{\omega }_{1}\times \hat{\mathbf{n}}_{1})\cdot (\mathbf{v}_{Q}-%
\mathbf{v}_{P})+\hat{\mathbf{n}}_{1}\cdot (\dot{\mathbf{v}}_{Q}-\dot{\mathbf{%
v}}_{P})+\mathbf{(\mathbf{v}}_{Q}-\mathbf{\mathbf{v}}_{P}\mathbf{)}\cdot
\mathbf{H}_{1}\cdot (\mathbf{v}_{Q}-\mathbf{v}_{P})  \notag \\
&=&-\hat{\mathbf{n}}_{1}\times (\bm{\omega }_{1}+\bm{\omega }_{2})\cdot (%
\mathbf{v}_{Q}-\mathbf{v}_{P})+\hat{\mathbf{n}}_{1}\cdot (\dot{\mathbf{v}}%
_{Q}-\dot{\mathbf{v}}_{P})  \notag \\
&&+\hat{\mathbf{n}}_{1}\times (\bm{\omega }_{2}-\bm{\omega }_{1})\cdot (%
\mathbf{v}_{Q}-\mathbf{v}_{P})+\mathbf{(\mathbf{v}}_{Q}-\mathbf{\mathbf{v}}%
_{P}\mathbf{)}\cdot \mathbf{H}_{1}\cdot (\mathbf{v}_{Q}-\mathbf{v}_{P}), \\
x_{2\varepsilon Q} &=&\frac{1}{2}(\hat{\bm{\varepsilon}}_{\bot }\cdot
\mathbf{H}_{2}\cdot \hat{\bm{\varepsilon}}_{\bot })+\frac{1}{2}(\hat{%
\bm{\varepsilon}}_{\bot }\cdot \mathbf{H}_{1}\cdot \hat{\bm{\varepsilon}}%
_{\bot })=\frac{1}{2}\hat{\bm{\varepsilon}}_{\bot }\left( \mathbf{H}_{2}+%
\mathbf{H}_{1}\right) \hat{\bm{\varepsilon}}_{\bot }, \\
v_{\varepsilon Q} &=&\hat{\mathbf{n}}_{1}\times (\bm{\omega }_{2}-\bm{\omega
}_{1})\cdot \hat{\bm{\varepsilon}}\bot +(\mathbf{v}_{Q}-\mathbf{v}_{P})\cdot
\mathbf{H}_{1}\cdot \hat{\bm{\varepsilon}}_{\bot }.
\end{eqnarray}%
The maximum of $\frac{v_{\varepsilon Q}^{2}}{2x_{2\varepsilon Q}}$ is
obtained at the direction
\begin{equation}
\hat{\bm{\varepsilon}}_{\bot Q}^{\ast }=\frac{\left( \mathbf{H}_{2}+\mathbf{H%
}_{1}\right) ^{-1}\cdot (\mathbf{\hat{n}}_{1}\times (\bm{\omega }_{2}-%
\bm{\omega }_{1})+\mathbf{H}_{1}\cdot (\mathbf{v}_{Q}-\mathbf{v}_{P}))}{%
|\left( \mathbf{H}_{2}+\mathbf{H}_{1}\right) ^{-1}\cdot (\hat{\mathbf{n}}%
_{1}\times (\bm{\omega }_{2}-\bm{\omega }_{1})+\mathbf{H}_{1}\cdot (\mathbf{v%
}_{Q}-\mathbf{v}_{P}))|}.
\end{equation}%
To make our presentation simpler, we denote
\begin{equation}
L_{Q^{\ast }}=|\left( \mathbf{H}_{2}+\mathbf{H}_{1}\right) ^{-1}\cdot (\hat{%
\mathbf{n}}_{1}\times (\bm{\omega }_{2}-\bm{\omega }_{1})+\mathbf{H}%
_{1}\cdot (\mathbf{v}_{Q}-\mathbf{v}_{P}))|.
\end{equation}%
Then direct substitution of $\hat{\bm{\varepsilon}}_{\bot Q}^{\ast }$ gives
\begin{equation}
v_{\varepsilon Q}^{\ast }=(\hat{\mathbf{n}}_{1}\times (\bm{\omega }_{2}-%
\bm{\omega }_{1})+(\mathbf{v}_{Q}-\mathbf{v}_{P})\cdot \mathbf{H}_{1})\cdot
\hat{\bm{\varepsilon}}_{\bot Q}^{\ast }=L_{Q^{\ast }}2x_{2\varepsilon
Q}^{\ast },
\end{equation}%
and this far in this part%
\begin{eqnarray}
&&(2a_{Q}x_{2\varepsilon Q}-v_{\varepsilon Q}^{2})(\hat{\bm{\varepsilon}}%
_{\bot }^{\ast })  \notag \\
&=&(2(\bm{\omega }_{1}\times \hat{\mathbf{n}}_{1})\cdot (\mathbf{v}_{Q}-%
\mathbf{v}_{P})+\hat{\mathbf{n}}_{1}\cdot (\dot{\mathbf{v}}_{Q}-\dot{\mathbf{%
v}}_{P})+\mathbf{(\mathbf{v}}_{Q}-\mathbf{\mathbf{v}}_{P}\mathbf{)}\cdot
\mathbf{H}_{1}\cdot (\mathbf{v}_{Q}-\mathbf{v}_{P}))v_{\varepsilon Q}^{\ast
}/L_{Q^{\ast }}-v_{\varepsilon Q}^{\ast }{}^{2}  \notag \\
&=&v_{\varepsilon Q}^{\ast }/L_{Q^{\ast }}[(-(\hat{\mathbf{n}}_{1}\times (%
\bm{\omega }_{1}+\bm{\omega }_{2}))\cdot (\mathbf{v}_{Q}-\mathbf{v}_{P})+%
\hat{\mathbf{n}}_{1}\cdot (\dot{\mathbf{v}}_{Q}-\dot{\mathbf{v}}_{P})  \notag
\\
&&+\hat{\mathbf{n}}_{1}\times (\bm{\omega }_{2}-\bm{\omega }_{1})\cdot (%
\mathbf{v}_{Q}-\mathbf{v}_{P})+\mathbf{(\mathbf{v}}_{Q}-\mathbf{\mathbf{v}}%
_{P}\mathbf{)}\cdot \mathbf{H}_{1}\cdot (\mathbf{v}_{Q}-\mathbf{v}%
_{P}))-v_{\varepsilon Q}^{\ast }{}L_{Q^{\ast }}]  \notag \\
&=&v_{\varepsilon Q}^{\ast }/L_{Q^{\ast }}[(-(\hat{\mathbf{n}}_{1}\times (%
\bm{\omega }_{1}+\bm{\omega }_{2}))\cdot (\mathbf{v}_{Q}-\mathbf{v}_{P})+%
\hat{\mathbf{n}}_{1}\cdot (\dot{\mathbf{v}}_{Q}-\dot{\mathbf{v}}_{P})]
\notag \\
&&+v_{\varepsilon Q}^{\ast }/L_{Q^{\ast }}[(\hat{\mathbf{n}}_{1}\times (%
\bm{\omega }_{2}-\bm{\omega }_{1})+\mathbf{H}_{1}\cdot (\mathbf{v}_{Q}-%
\mathbf{v}_{P}))  \notag \\
&&\cdot ((\mathbf{v}_{Q}-\mathbf{v}_{P})-\left( \mathbf{H}_{2}+\mathbf{H}%
_{1}\right) ^{-1}\cdot (\hat{\mathbf{n}}_{1}\times (\bm{\omega }_{2}-%
\bm{\omega }_{1})+\mathbf{H}_{1}\cdot (\mathbf{v}_{Q}-\mathbf{v}_{P}))]
\notag \\
&=&2x_{2\varepsilon Q}^{\ast }[(-(\hat{\mathbf{n}}_{1}\times (\bm{\omega }%
_{1}+\bm{\omega }_{2}))\cdot (\mathbf{v}_{Q}-\mathbf{v}_{P})+\hat{\mathbf{n}}%
_{1}\cdot (\dot{\mathbf{v}}_{Q}-\dot{\mathbf{v}}_{P})  \notag \\
&&+(\hat{\mathbf{n}}_{1}\times (\bm{\omega }_{2}-\bm{\omega }_{1})+\mathbf{H}%
_{1}\cdot (\mathbf{v}_{Q}-\mathbf{v}_{P}))  \notag \\
&&\cdot ((\mathbf{v}_{Q}-\mathbf{v}_{P})-\left( \mathbf{H}_{2}+\mathbf{H}%
_{1}\right) ^{-1}\cdot (\hat{\mathbf{n}}_{1}\times (\bm{\omega }_{2}-%
\bm{\omega }_{1})+\mathbf{H}_{1}\cdot (\mathbf{v}_{Q}-\mathbf{v}_{P}))].
\end{eqnarray}

Similarly, for potentially most deeply penetrating neighboring point on body
1, we have%
\begin{equation}
F_{2}(\mathbf{r}_{P+}(t),t)=v_{P}t+\frac{1}{2}(a_{P}-\frac{v_{\varepsilon
P}^{2}}{x_{2\varepsilon P}})t^{2}+O_{\varepsilon }(t^{3}),
\end{equation}%
where%
\begin{eqnarray}
v_{P} &=&\hat{\mathbf{n}}_{2}\cdot (\mathbf{v}_{P}-\mathbf{v}_{Q}), \\
a_{P} &=&2(\bm{\omega }_{2}\times \hat{\mathbf{n}}_{2})\cdot (\mathbf{v}_{P}-%
\mathbf{v}_{Q})+\hat{\mathbf{n}}_{2}\cdot (\dot{\mathbf{v}}_{P}-\dot{\mathbf{%
v}}_{Q})+\mathbf{(\mathbf{v}}_{P}-\mathbf{\mathbf{v}}_{Q}\mathbf{)}\cdot
\mathbf{H}_{2}\cdot (\mathbf{v}_{P}-\mathbf{v}_{Q}), \\
x_{2\varepsilon P} &=&\frac{1}{2}(\hat{\bm{\varepsilon}}_{\bot }\cdot
\mathbf{H}_{1}\cdot \hat{\bm{\varepsilon}}_{\bot })+\frac{1}{2}(\hat{%
\bm{\varepsilon}}_{\bot }\cdot \mathbf{H}_{2}\cdot \hat{\bm{\varepsilon}}%
_{\bot }), \\
v_{\varepsilon P} &=&\hat{\mathbf{n}}_{2}\times (\bm{\omega }_{1}-\bm{\omega
}_{2})\cdot \hat{\bm{\varepsilon}}_{\bot }+(\mathbf{v}_{P}-\mathbf{v}%
_{Q})\cdot \mathbf{H}_{2}\cdot \hat{\bm{\varepsilon}_{\bot },}
\end{eqnarray}%
The maximum of $\frac{v_{\varepsilon P}^{2}}{2x_{2\varepsilon P}}$ is
obtained in the direction
\begin{equation}
\hat{\bm{\varepsilon}}_{\bot P}^{\ast }=\frac{\left( \mathbf{H}_{2}+\mathbf{H%
}_{1}\right) ^{-1}\cdot (\mathbf{\hat{n}}_{2}\times (\bm{\omega }_{1}-%
\bm{\omega }_{2})+\mathbf{H}_{2}\cdot (\mathbf{v}_{P}-\mathbf{v}_{Q}))}{%
|\left( \mathbf{H}_{2}+\mathbf{H}_{1}\right) ^{-1}\cdot (\mathbf{\hat{n}}%
_{2}\times (\bm{\omega }_{1}-\bm{\omega }_{2})+\mathbf{H}_{2}\cdot (\mathbf{v%
}_{P}-\mathbf{v}_{Q}))|}.
\end{equation}%
We denote
\begin{equation}
L_{P^{\ast }}=|(\left( \mathbf{H}_{2}+\mathbf{H}_{1}\right) ^{-1}\cdot (\hat{%
\mathbf{n}}_{2}\times (\bm{\omega }_{1}-\bm{\omega }_{2})+\mathbf{H}%
_{2}\cdot (\mathbf{v}_{P}-\mathbf{v}_{Q}))|,
\end{equation}%
\begin{equation}
v_{\varepsilon P}^{\ast }=(\hat{\mathbf{n}}_{2}\times (\bm{\omega }_{1}-%
\bm{\omega }_{2})+(\mathbf{v}_{P}-\mathbf{v}_{Q})\cdot \mathbf{H}_{2})\cdot
\hat{\bm{\varepsilon}}_{\bot P}^{\ast }=L_{P^{\ast }}2x_{2\varepsilon
P}^{\ast }.
\end{equation}%
Using the fact that%
\begin{equation}
\hat{\mathbf{n}}_{2}=-\hat{\mathbf{n}}_{1},
\end{equation}%
we have%
\begin{eqnarray}
&&(2a_{P}x_{2\varepsilon P}-v_{\varepsilon P}^{2})(\hat{\bm{\varepsilon}}%
_{\bot }^{\ast })  \notag \\
&=&v_{\varepsilon P}^{\ast }/L_{P^{\ast }}[((\hat{\mathbf{n}}_{2}\times (%
\bm{\omega }_{1}+\bm{\omega }_{2}))\cdot (\mathbf{v}_{Q}-\mathbf{v}_{P})-%
\hat{\mathbf{n}}_{2}\cdot (\dot{\mathbf{v}}_{Q}-\dot{\mathbf{v}}_{P})]
\notag \\
&&+v_{\varepsilon P}^{\ast }/L_{P^{\ast }}[(-\hat{\mathbf{n}}_{2}\times (%
\bm{\omega }_{2}-\bm{\omega }_{1})-\mathbf{H}_{2}\cdot (\mathbf{v}_{Q}-%
\mathbf{v}_{P}))  \notag \\
&&\cdot (-(\mathbf{v}_{Q}-\mathbf{v}_{P})-\left( \mathbf{H}_{2}+\mathbf{H}%
_{1}\right) ^{-1}\cdot (-\hat{\mathbf{n}}_{2}\times (\bm{\omega }_{2}-%
\bm{\omega }_{1})-\mathbf{H}_{2}\cdot (\mathbf{v}_{Q}-\mathbf{v}_{P}))]
\notag \\
&=&v_{\varepsilon P}^{\ast }/L_{P^{\ast }}[((-\hat{\mathbf{n}}_{1}\times (%
\bm{\omega }_{1}+\bm{\omega }_{2}))\cdot (\mathbf{v}_{Q}-\mathbf{v}_{P})+%
\hat{\mathbf{n}}_{1}\cdot (\dot{\mathbf{v}}_{Q}-\dot{\mathbf{v}}_{P})]
\notag \\
&&+v_{\varepsilon P}^{\ast }/L_{P^{\ast }}[(\hat{\mathbf{n}}_{1}\times (%
\bm{\omega }_{2}-\bm{\omega }_{1})-\mathbf{H}_{2}\cdot (\mathbf{v}_{Q}-%
\mathbf{v}_{P}))  \notag \\
&&\cdot (-(\mathbf{v}_{Q}-\mathbf{v}_{P})-\left( \mathbf{H}_{2}+\mathbf{H}%
_{1}\right) ^{-1}\cdot (\hat{\mathbf{n}}_{1}\times (\bm{\omega }_{2}-%
\bm{\omega }_{1})-\mathbf{H}_{2}\cdot (\mathbf{v}_{Q}-\mathbf{v}_{P}))]
\notag \\
&=&2x_{2\varepsilon P}^{\ast }[((-\hat{\mathbf{n}}_{1}\times (\bm{\omega }%
_{1}+\bm{\omega }_{2}))\cdot (\mathbf{v}_{Q}-\mathbf{v}_{P})+\hat{\mathbf{n}}%
_{1}\cdot (\dot{\mathbf{v}}_{Q}-\dot{\mathbf{v}}_{P})  \notag \\
&&+(\hat{\mathbf{n}}_{1}\times (\bm{\omega }_{2}-\bm{\omega }_{1})-\mathbf{H}%
_{2}\cdot (\mathbf{v}_{Q}-\mathbf{v}_{P}))  \notag \\
&&\cdot (-(\mathbf{v}_{Q}-\mathbf{v}_{P})-\left( \mathbf{H}_{2}+\mathbf{H}%
_{1}\right) ^{-1}\cdot (\hat{\mathbf{n}}_{1}\times (\bm{\omega }_{2}-%
\bm{\omega }_{1})-\mathbf{H}_{2}\cdot (\mathbf{v}_{Q}-\mathbf{v}_{P}))].
\end{eqnarray}%
Using the fact that%
\begin{equation}
\mathbf{H}_{2}\cdot \left( \mathbf{H}_{2}+\mathbf{H}_{1}\right) ^{-1}=%
\mathbf{I}-\mathbf{H}_{1}\cdot \left( \mathbf{H}_{2}+\mathbf{H}_{1}\right)
^{-1},
\end{equation}%
the above expression, can be shown to be, via straightforward calculation,%
\begin{eqnarray}
&&(2a_{P}x_{2\varepsilon P}-v_{\varepsilon P}^{2})(\hat{\bm{\varepsilon}}%
_{\bot }^{\ast })  \notag \\
&=&2x_{2\varepsilon P}^{\ast }[((-\hat{\mathbf{n}}_{1}\times (\bm{\omega }%
_{1}+\bm{\omega }_{2}))\cdot (\mathbf{v}_{Q}-\mathbf{v}_{P})+\hat{\mathbf{n}}%
_{1}\cdot (\dot{\mathbf{v}}_{Q}-\dot{\mathbf{v}}_{P})  \notag \\
&&+(\hat{\mathbf{n}}_{1}\times (\bm{\omega }_{2}-\bm{\omega }_{1})+\mathbf{H}%
_{1}\cdot (\mathbf{v}_{Q}-\mathbf{v}_{P}))  \notag \\
&&\cdot ((\mathbf{v}_{Q}-\mathbf{v}_{P})-\left( \mathbf{H}_{2}+\mathbf{H}%
_{1}\right) ^{-1}\cdot (\hat{\mathbf{n}}_{1}\times (\bm{\omega }_{2}-%
\bm{\omega }_{1})+\mathbf{H}_{1}\cdot (\mathbf{v}_{Q}-\mathbf{v}_{P})).
\end{eqnarray}

Thus%
\begin{equation}
\frac{2a_{P}x_{2\varepsilon P}^{\ast }-v_{\varepsilon P}^{\ast 2}}{%
2x_{2\varepsilon P}^{\ast }}=\frac{2a_{Q}x_{2\varepsilon Q}^{\ast
}-v_{\varepsilon Q}^{\ast 2}}{x_{2\varepsilon Q}^{\ast }},
\end{equation}%
or%
\begin{equation}
a_{P}-(\frac{v_{\varepsilon P}^{2}}{2x_{2\varepsilon P}})_{\max }=a_{Q}-(%
\frac{v_{\varepsilon Q}^{2}}{2x_{2\varepsilon Q}})_{\max },
\end{equation}%
which indicates that the terms $a_{P}-(\frac{v_{\varepsilon P}^{2}}{%
2x_{2\varepsilon P}})_{\max }$ and $a_{Q}-(\frac{v_{\varepsilon Q}^{2}}{%
2x_{2\varepsilon Q}})_{\max }$ will be positive, zero, or negative
simutaneously.

\subsubsection{Proof of $\mathbf{a}_{P}\cdot \mathbf{n<0}$ for a uniform
ellipsoid}

Here we compute the normal acceleration of a point on the boundary of a
freely moving ellipsoidal body. The equation of the surface is given by%
\begin{equation}
\mathbf{p}\cdot \mathbf{A}\cdot\mathbf{\ p}=1,
\end{equation}%
where $\mathbf{p}=p_{x}\hat{\mathbf{x}}+p_{y}\hat{\mathbf{y}}+p_{z}\mathbf{%
\hat{z}}$ is the vector from the center of mass to a point $\mathbf{p}$ on
the surface, and%
\begin{equation}
\mathbf{A}=\left[
\begin{array}{lll}
\frac{1}{a^{2}} & 0 & 0 \\
0 & \frac{1}{b^{2}} & 0 \\
0 & 0 & \frac{1}{c^{2}}%
\end{array}%
\right] ,
\end{equation}%
where $a,b,c$ are the semi-major axes lengths.

The acceleration of a point on the surface of the ellipsoid is given by%
\begin{equation}
\mathbf{a}_{P}=\dot{\bm{\omega}}\times \mathbf{p}+\bm{\omega} (\bm{\omega}
\cdot \mathbf{p} )-\omega ^{2}\mathbf{p},
\end{equation}%
where
\begin{equation}
\dot{\bm{\omega}}=-\mathbf{I}^{-1}\cdot (\bm{\omega} \times \mathbf{I}\cdot %
\bm{\omega} )\times \mathbf{p}.
\end{equation}

It follows at once that the component of the acceleration $\mathbf{a}_{P}$
in the $\hat{\mathbf{p}}$ direction is non-positive; that is%
\begin{equation}
\mathbf{a}_{P}\cdot \hat{\mathbf{p}}=\omega ^{2} p((\hat{\bm{\omega}}\cdot
\hat{\mathbf{p}})^{2}-1)\leq 0.
\end{equation}

We write%
\begin{equation}
\mathbf{I}=\left[
\begin{array}{ccc}
\alpha & 0 & 0 \\
0 & \beta & 0 \\
0 & 0 & \gamma%
\end{array}%
\right]
\end{equation}%
and%
\begin{equation}
\mathbf{I}^{-1}=\left[
\begin{array}{ccc}
\frac{1}{\alpha } & 0 & 0 \\
0 & \frac{1}{\beta } & 0 \\
0 & 0 & \frac{1}{\gamma }%
\end{array}%
\right] .
\end{equation}%
For an uniform ellipsoid
\begin{eqnarray}
\alpha &=&\frac{1}{5}M(b^{2}+c^{2}), \\
\beta &=&\frac{1}{5}M(a^{2}+c^{2}), \\
\gamma &=&\frac{1}{5}M(a^{2}+b^{2}).
\end{eqnarray}

A direct evaluation of $\dot{\bm{\omega}}=-\mathbf{I}^{-1}\cdot (\bm{\omega}%
\times \mathbf{I}\cdot \bm{\omega})$ gives%
\begin{eqnarray}
\dot{\omega}_{x} &=&\omega _{y}\omega _{z}(\frac{c^{2}-b^{2}}{c^{2}+b^{2}}),
\\
\dot{\omega}_{y} &=&\omega _{x}\omega _{z}(\frac{a^{2}-c^{2}}{a^{2}+c^{2}}),
\\
\dot{\omega}_{z} &=&\omega _{x}\omega _{y}(\frac{b^{2}-a^{2}}{b^{2}+a^{2}}),
\end{eqnarray}%
We note that the normal $\mathbf{n}$ of the ellipsoid at point $P$ is given
by%
\begin{equation}
\mathbf{n}=\frac{p_{x}}{a^{2}}\hat{\mathbf{x}}+\frac{p_{y}}{b^{2}}\mathbf{%
\hat{y}+}\frac{p_{z}}{c^{2}}\hat{\mathbf{z}}.
\end{equation}%
We can write the normal acceleration as%
\begin{eqnarray}
\mathbf{a}_{P}\cdot \mathbf{n} &=&(\dot{\bm{\omega}}\times p)\cdot \mathbf{n}%
+\bm{\omega }(\bm{\omega }\cdot \mathbf{p})\cdot \mathbf{n}-\omega ^{2}%
\mathbf{p\cdot n}  \notag \\
&=&(\dot{\omega}_{y}p_{z}-\dot{\omega}_{z}p_{y}\mathbf{)}\frac{p_{x}}{a^{2}}%
\mathbf{+}(\dot{\omega}_{z}p_{x}-\dot{\omega}_{x}p_{z})\frac{p_{y}}{b^{2}}%
\mathbf{+(}\dot{\omega}_{x}p_{y}-\dot{\omega}_{y}p_{x}\mathbf{)}\frac{p_{z}}{%
c^{2}}  \notag \\
&&+(\omega _{x}p_{x}+\omega _{y}p_{y}+\omega _{z}p_{z}\mathbf{)(}\omega _{x}%
\frac{p_{x}}{a^{2}}\mathbf{+}\omega _{y}\frac{p_{y}}{b^{2}}\mathbf{+}\omega
_{z}\frac{p_{z}}{c^{2}})  \notag \\
&&-(\omega _{x}^{2}+\omega _{y}^{2}+\omega _{z}^{2})(\frac{p_{x}^{2}}{a^{2}}%
\mathbf{+}\frac{p_{y}^{2}}{b^{2}}\mathbf{+}\frac{p_{z}^{2}}{c^{2}}).
\end{eqnarray}%
Explicitly%
\begin{eqnarray}
\mathbf{a}_{P}\cdot \mathbf{n} &=&\omega _{x}\omega _{y}p_{x}p_{y}((\frac{1}{%
b^{2}}-\frac{1}{a^{2}})\frac{b^{2}-a^{2}}{b^{2}+a^{2}}+(\frac{1}{b^{2}}+%
\frac{1}{a^{2}}))  \notag \\
&&+\omega _{x}\omega _{z}p_{x}p_{z}((\frac{1}{a^{2}}-\frac{1}{c^{2}})\frac{%
a^{2}-c^{2}}{a^{2}+c^{2}}+(\frac{1}{a^{2}}+\frac{1}{c^{2}}))  \notag \\
&&+\omega _{y}\omega _{z}p_{y}p_{z}((\frac{1}{c^{2}}-\frac{1}{b^{2}})\frac{%
c^{2}-b^{2}}{c^{2}+b^{2}}+(\frac{1}{a^{2}}+\frac{1}{c^{2}}))  \notag \\
&&+\omega _{x}^{2}p_{x}^{2}\frac{1}{a^{2}}+\omega _{y}^{2}p_{y}^{2}\frac{1}{%
b^{2}}+\omega _{z}^{2}p_{z}^{2}\frac{1}{c^{2}}  \notag \\
&&-(\omega _{x}^{2}+\omega _{y}^{2}+\omega _{z}^{2}).
\end{eqnarray}%
We note that%
\begin{equation}
(\frac{1}{b^{2}}-\frac{1}{a^{2}})\frac{b^{2}-a^{2}}{b^{2}+a^{2}}+(\frac{1}{%
b^{2}}+\frac{1}{a^{2}})=\frac{4}{(a^{2}+b^{2})}\leq \frac{2}{ab},
\end{equation}%
thus%
\begin{eqnarray}
\mathbf{a}_{P}\cdot \mathbf{n} &\leq &(\frac{\omega _{x}p_{x}}{a}+\frac{%
\omega _{y}p_{y}}{b}+\frac{\omega _{z}p_{z}}{c})^{2}-(\omega _{x}^{2}+\omega
_{y}^{2}+\omega _{z}^{2})  \notag \\
&=&(\bm{\omega }\cdot \mathbf{A}^{1/2}\cdot \mathbf{p})^{2}-\omega ^{2}\leq
0,
\end{eqnarray}%
the last inequality holds since $(\mathbf{A}^{1/2}\cdot \mathbf{p})^{2}=%
\mathbf{p\cdot A\cdot p=}1$, and so $\mathbf{A}^{1/2}\cdot \mathbf{p}$ is a
unit vector.

\end{document}